\documentclass[aps,pre,superscriptaddress,twocolumn,eqsecnum,floatfix,showpacs]{revtex4}
\usepackage{amsmath,graphicx,txfonts}

\newcommand{\pa}[2]{\partial #1/\partial #2}
\newcommand{\im}{\mathrm{i}}
\newcommand{\br}{\mathbf{r}}
\begin{document}
\title{Financial Atoms and Molecules}
\author{Yik Wen \surname{GOO}}
\thanks{These authors contributed equally to this work.}
\affiliation{CN Yang Scholars Programme, Nanyang Technological
University}
\affiliation{School of Mechanical and Aerospace Engineering,
Nanyang Technological University, 50 Nanyang Avenue, Singapore 639798}
\author{Tong Wei \surname{LIAN}}
\thanks{These authors contributed equally to this work.}
\affiliation{CN Yang Scholars Programme, Nanyang Technological
University}
\affiliation{School of Electronic and Electrical Engineering,
Nanyang Technological University, 50 Nanyang Avenue, Singapore 639798}
\author{Wei Guang \surname{ONG}}
\affiliation{CN Yang Scholars Programme, Nanyang Technological
University}
\affiliation{Division of Physics and Applied Physics, School of
Physical and Mathematical Sciences, Nanyang Technological
University, 21 Nanyang Link, Singapore 637371}
\author{Wen Ting \surname{CHOI}}
\affiliation{Division of Physics and Applied Physics, School of
Physical and Mathematical Sciences, Nanyang Technological
University, 21 Nanyang Link, Singapore 637371}
\author{Siew Ann \surname{CHEONG}}
\thanks{Corresponding author.}
\email{cheongsa@ntu.edu.sg}
\affiliation{Division of Physics and Applied Physics, School of
Physical and Mathematical Sciences, Nanyang Technological
University, 21 Nanyang Link, Singapore 637371}
\date{\today}

\begin{abstract}
Atoms and molecules are important conceptual entities we invented to
understand the physical world around us.  The key to their usefulness
lies in the organization of nuclear and electronic degrees of freedom
into a single dynamical variable whose time evolution we can better
imagine.  The use of such effective variables in place of the true
microscopic variables is possible because of the separation between
nuclear time scales (very fast), electronic time scales (fast), atomic
time scales (slow), and molecular time scales (slower still).  Where
separation of time scales occurs, identification of analogous objects
in complex systems --- an example of which is the financial market ---
can help advance our understanding of their dynamics.  To detect
separated time scales and identify their associated effective degrees
of freedom in financial markets, we devised a two-stage statistical
clustering scheme to analyze the price movements of stocks in several
equity markets.  The price movements are first clustered at a short
time scale, and thereafter, re-clustered based on separated
correlation levels that become evident when these short-time
correlations are examined over a longer time scale.  Through this
two-time-scale clustering analysis, we discovered a hierarchy of
levels of self-organization in real financial markets, whereby lower
level, rapidly-evolving dynamical structures are nested within higher
level, slowly-evolving dynamical structures, which are themselves
nested within even higher level, even more slowly-evolving dynamical
structures.  We call these self-organized dynamical structures (which
are statistically robust) \emph{financial atoms}, \emph{financial
molecules}, and \emph{financial supermolecules}.  While many financial
atoms have compositions that on hindsight appear trivial or easy to
guess, there are also those whose composition are genuine surprises.
With larger and better-regulated markets, the component stocks of
financial molecules and financial supermolecules all fall within a
single market sector, or several very closely-related market sectors.
In markets with extensive cross ownership, or dominated by huge
multi-industry holding companies, financial molecules and financial
supermolecules straddling many distant market sectors were found.  In
general, the detailed compositions of these dynamical structures
cannot be deduced based on raw financial intuition alone, and must be
explained in terms of the underlying portfolios, and investment
strategies of market players.  More interestingly, we find that major
market events such as the \emph{Chinese Correction} and the
\emph{Subprime Crisis} leave many tell-tale signs within the
correlational structures of financial molecules.  

\end{abstract}


\maketitle

\section{Introduction}

Predicting the movement of prices in financial markets is an important
problem, more so in the light of the global financial crisis we find
ourselves in right now.  If we have the means to detect market
undercurrents that do not have clear signatures on the surface, we
might be able to steer clear of trouble, and perhaps even defuse
financial landmines before things get out of hand.  However, there are
at present no reliable schemes for short- or long-term predictions,
despite the fact that past information on all financial instruments
are wholly available \cite{Transactions}.  This stands in stark
contrast to our ability to predict solar and lunar eclipses to very
high precision, starting from Newton's Laws of Motion, and much less
voluminous initial data.  We believe this is due to the fact that
financial markets are complex systems, whose dynamics follow
self-organizing principles we do not yet understand.  Clearly, an
important first step towards writing down any predictive models of
financial market dynamics would be to identify the important degrees
of freedom.  \emph{A priori}, we have no clue whether these are
individual financial instruments, or specific collections of
instruments.  We hence look to the physical sciences for inspiration.

\subsection{Statistical Learning in the Physical Sciences}

In the physical sciences, we talk about atoms and molecules as if they
are real entities.  Quantum-mechanically, all atomic and molecular
systems must be described by a single many-body wave function
$\Psi(\br_1, \dots, \br_N, t)$ whose dynamics is governed by the
Schr\"odinger equation $\im\hbar(\pa{\Psi}{t}) = H\Psi$.  If we admit
that the wave function $\Psi(\br_1, \dots, \br_N, t)$ furnishes a
sufficient description of the problem, then atoms have no real
existence inside a molecule, and molecules have no real existence
within an interacting collection of molecules.  Nevertheless, it is
very useful to continue speaking of them, because they simplify our
mental models of the processes that take place within the collection
of nucleons and electrons.  This is admissible, and in fact physically
meaningful, because the dynamical structures that atoms and molecules
participate in evolve over time scales significantly longer than the
intrinsic time scales set by nuclear and electronic motion, i.e.
there is a separation between nuclear/electronic time scales and
atomic/molecular time scales.

To better illustrate how our association of certain collections of
nuclear and electronic degrees of freedom with atoms, and other
collections of nuclear and electronic degrees of freedom with
molecules arise naturally from this separation of time scales, let us
imagine a universe containing a single water molecule
$^1$H$_2$$^{16}$O, consisting of ten protons, eight neutrons, and ten
electrons.  Under ordinary conditions, the $^{16}$O nucleons and the
two $^1$H protons are always very well separated spatially, so in
principle we have no problem telling them apart.  For the ten
indistinguishable electrons, however, we cannot simply assign eight
labeled electrons to $^{16}$O, and one labeled electron each to the
two $^1$H.  Nevertheless, if we make a movie of the time evolution of
electronic densities within the water molecule, we will find an
electronic distribution close to $^{16}$O nucleus that evolves pretty
much independently of the electronic distributions between the
$^{16}$O nucleus and $^1$H nuclei.  We think of the former as the
density of $^{16}$O core electrons, and the latter as the density of
the bonding electrons.  Unlike the $^{16}$O core electrons, the
bonding electrons cannot be associated with either the $^{16}$O atom
or the $^1$H atom alone.  Instead, they must be associated with the
two chemical bonds --- another conceptual tool we invent to help us
understand the nature of stable bound states between atomic species
--- in the water molecule.

Now let us imagine actually producing such a movie, through direct
integration of the time-dependent Schr\"odinger equation for the water
molecule.  If we assume that the nucleons are static, we can solve
this problem in the basis of all possible 10-electron configurations.
Some of these configurations will contribute predominantly to the core
density, while others will contribute predominantly to the bond
density.  Because of the nucleon-electron and electron-electron
Coulomb interactions, the quantum-mechanical amplitudes of these
configurations will evolve with time, sometimes rapidly, and sometimes
slowly.  As a result of these variable rates of time evolution, if we
examine the configurations contributing predominantly to the $^{16}$O
core electronic density, over a time interval that is on the order of
the electronic time scale, we might find their amplitudes to be
sometimes correlated, and sometimes uncorrelated.  In fact, over
electronic time scales, some core electronic amplitudes might even be
more correlated with bonding electronic amplitudes than with other
core electronic amplitudes.  Nevertheless, based on our intuitive
picture of distinguishable core and bonding electronic densities, and
the separation of electronic and atomic/molecular time scales, we
expect the long-run averages of core-core short-time correlations, as
well as the long-run averages of bond-bond short-time correlations, to
tend towards some nonzero values, while the long-run averages of
core-bonding correlations tend towards zero.  

\subsection{Atoms and Molecules as Robust Dynamical Structures}

The understanding that emerges from this dynamical picture is that
core and bond amplitudes should not be treated as objects with
immutable labels.  Instead, they ought to be identified as clusters
emerging from a statistical learning procedure, where we repeatedly
calculate correlations at the electronic time scale (the short time
scale), and examine how the patterns of short-time correlations look
like over a much longer duration (the long time scale).   Given a
collection of unlabeled time-dependent amplitudes, we can always make
use of this procedure of statistical clustering at two different
(separated) time scales to discover the compound objects, things that
we would go on to call \emph{atoms} and \emph{molecules}.  More
generally, if we are given a large number of time-dependent variables,
determining how short-time correlations are themselves correlated over
longer times would allow us to determine which sets of variables are
strongly correlated over long time scales.  These strongly-correlated
set of variables then serve as good candidates for effective degrees
of freedom at the long time scale.

\begin{figure}[htbp]
\centering
\includegraphics[width=\linewidth]{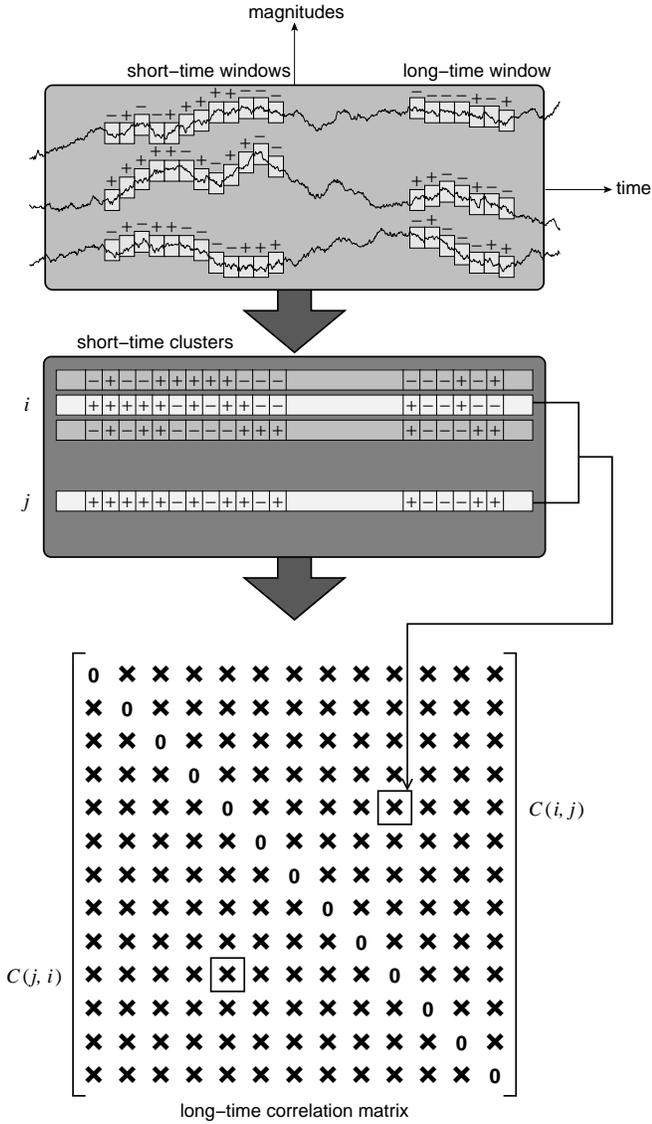}
\caption{Scheme of the two-time-scale statistical clustering method.
In the top panel, the time series of $N$ variables are examined within
a long-time window, which consists of a large number of short-time
windows.  In short-time window $t$, variable $i$ is assigned to
the `$+$' cluster if $\Delta x_i(t) = x_i(t+1) - x_i(t) > 0$, and to
the `$-$' cluster if $\Delta x_i(t) = x_i(t+1) - x_i(t) < 0$, as shown
in the middle panel.  In the bottom panel, we compute long-time correlation
matrix $C$, such that $C(i, j)$ is the number of times variables $i$
and $j$ are assigned to the \emph{same} short-time cluster.}
\label{fig:twotimescale}
\end{figure}

Ultimately, for dynamical features as robust as the core electronic
densities and the bonding electronic densities, we expect their
statistical signatures to be generally insensitive to specific choices of
the short and long time scales, so long as the short time scale is
comparable to electronic time scales, and the long time scale is
comparable to the atomic/molecular time scale.  In addition, it should
not matter what short-time correlations we evaluate, and what
long-time correlations of these we further determine.  For some
choices of short-time correlations, statistical signatures of the core
and bond clusters of amplitudes might emerge fairly quickly during the
long-time clustering.  For other choices of short-time correlations,
we might have to cluster over a longer time to detect these
statistical signatures.  Recast as a learning problem, the extraction
of robust, and therefore physically meaningful, dynamical features
boils down to a simple problem of statistics and statistical
significance.  The simplest scheme that implements the two-time-scale
statistical clustering procedure outlined above is based on the sign
of the rate of change, as shown in Figure \ref{fig:twotimescale}.  In
this scheme, two scalar variables $x_i$ and $x_j$ are correlated over
a short-time window if they are both increasing or decreasing.
Otherwise, the two variables are considered to be uncorrelated.  Over
the long time scale, we define their long-time correlation $C(i, j)$
to be the number of times they are correlated with each other at short
time scales.  These pairwise correlations can be organized into a
long-time correlation matrix $C$, which is the starting point for our
identification of strongly-correlated compound objects.

\section{Data and Results}

Armed with this understanding, we analyzed five financial markets, the
New York Stock Exchange (NYSE), the London Stock Exchange (LSE), the
Tokyo Stock Exchange (TSE), the Hong Kong Stock Exchange (HKSE), and
the Singapore Stock Exchange (SGX) over a two-year period from January
2006 to December 2007 (see Table \ref{tab:markets}).  These markets
are chosen specifically for contrast: the NYSE, LSE, and TSE are
mature markets in G7 countries, whereas the HKSE and SGX are emerging
markets.  Presumably, the dynamics in emerging markets should be
different from that in mature markets, and thus one might naively
expect structurally different self-organizations in these two classes
of markets.  To keep the datasets manageable, we restricted our
analysis to only stocks and stock related securities.  Since stocks
are important components in any financial market, we believe their
dynamics alone would offer us valuable insights into the inner
workings of the financial world.  

\begin{table}[htbp]
\centering
\caption{The total number of stocks and market sectors monitored from
January 2006 to December 2007, for each of the five equity markets
studied.  These five markets are chosen to contrast mature markets in
G7 countries (one each in North America, Europe, and Asia), and
emerging markets in Asia.}
\label{tab:markets}
\vskip .5\baselineskip
\begin{tabular}{lccccc}
\hline
& NYSE & LSE & TSE & HKSE & SGX \\
\hline
Total number of stocks & 6743\textsuperscript{1} &
8820\textsuperscript{2} & 2636\textsuperscript{2} &
1070\textsuperscript{3} & 546\textsuperscript{1} \\
Total number of market sectors & 10\textsuperscript{3} &
10\textsuperscript{5} & 39\textsuperscript{5} & 12\textsuperscript{4}
& 12\textsuperscript{6} \\
\hline
\multicolumn{6}{l}{\textsuperscript{1}Thomson Datastream,
\url{http://www.datastream.com/}.} \\
\multicolumn{6}{l}{\textsuperscript{2}Taqtic, SIRCA,
\url{https://taqtic.sirca.org.au/TaqTic/}.} \\
\multicolumn{6}{l}{\textsuperscript{3}\url{http://finance.yahoo.com/}
\cite{Renfree}.} \\
\multicolumn{6}{l}{\textsuperscript{4}Hong Kong Stock Exchange,} \\
\multicolumn{6}{l}{\url{http://www.hkex.com.hk/invest/index.asp?}} \\
\multicolumn{6}{l}{\url{id=company/profilemenu_page_e.asp}.} \\
\multicolumn{6}{l}{\textsuperscript{5}Reuters.} \\
\multicolumn{6}{l}{\textsuperscript{6}SGX - Singapore Exchange Ltd,} \\
\multicolumn{6}{l}{\url{http://info.sgx.com/webstocks.nsf/revamp+new+all+}} \\
\multicolumn{6}{l}{\url{stocks+sector?OpenView}.}
\end{tabular}
\end{table}

\subsection{Financial Atoms and Atomic Correlation Levels}

For each financial market, we looked at the daily price movements
$\{\Delta p_1(t), \Delta p_2(t), \dots, \Delta p_i(t), \dots, \Delta
p_N(t)\}$ of the $N$ stocks making up the market.  Here, $\Delta
p_i(t) = p_i(t+1) - p_i(t)$, and $p_i(t)$ is the price of stock $i$
on day $t$.  For this choice of data frequency, we have effectively
selected a short time scale of one day.  Our long time scale, chosen
to ensure statistical significance \cite{Significance}, is the
two-year observation period, consisting of 500+ short time windows.
It is useful to keep in mind that for all financial markets, there is
a microscopic time scale set by the average trading interval, which
can be anywhere between $10^{-2}$ s to $10^1$ s.  Following the
two-time-scale clustering procedure outlined above, we assign stocks
$i$ and $j$ to the same short-time cluster at time $t$, if $\Delta
p_i(t)$ and $\Delta p_j(t)$ are of the same sign.  A stock $k$ whose
price does not change from day $t$ to day $t+1$, i.e.  $\Delta p_k(t)
= 0$, is left unassigned.  In the long-time correlation matrix $C$
that emerges from this procedure, we find a background correlation
level $c_0$ reflecting market-level drifts, experienced by all stock
prices, that results from rallies and crashes.  This background
correlation level $c_0$ varies from market to market.  In general, the
larger $c_0$ is, the more liquid the market.

To get a rough sense of the dynamical structures having long-time
correlations over and above market-level drifts, we zero matrix
elements in $C$ below a threshold $c > c_0$.  As $c$ is raised, we go
from a dense matrix, telling us that each stock is correlated with a
large number of stocks (though not equally strongly), to a sparse
matrix, telling us that each stock is correlated only with very few
other stocks (but very strongly so).  Just as surfaces of high
constant electronic density frequently reveal the atomic constituents
of a molecule, very large matrix elements in $C$ reveal the atomic
constituents of the given financial market.  We call these small
clusters of very strongly correlated stocks \emph{financial atoms}.
The correlation levels within financial atoms rise significantly above
the background level $c_0$, because the dynamics of atomic stocks are
coherent over time scales much longer than the average trading
interval, which plays the role of the `nuclear/electronic' time scale.
The existence of such financial atoms is evident from earlier studies
\cite{Mantegna1999EurPhysJB11p193, Bonanno2001QuantFin1p96,
Onnela2002EurPhysJB30p285, Onnela2003PhysicaScriptaT106p48,
Onnela2003PhysRevE68e056110, Bonanno2004EurPhysJB38p363,
Onnela2004EurPhysJB38p353, Coronello2005ActaPhysPolB36p2653,
Hawkesby2005SSRN724062, Jung2005ProcSPIE5848,
Kim2005PhysRevE72e046133, Jung2006PhysicaA361p263,
Coelho2007PhysicaA373p615, Coronello2007ProcSPIE66010T, 
Heimo2007PhysicaA383p147, Pan2007PhysRevE76e046116,
Sinha2007EconophysMarketsBusNet, Zovko2007ProcComplexity}, where they
are referred to as \emph{synchronized clusters}.  However, the
separation of `nuclear/electronic' and atomic time scales was not
recognized.  The potential of using financial atoms as effective
variables to describe real financial markets is also not widely
appreciated.

\begin{figure}[htbp]
\centering
\includegraphics[scale=0.45, clip=true]{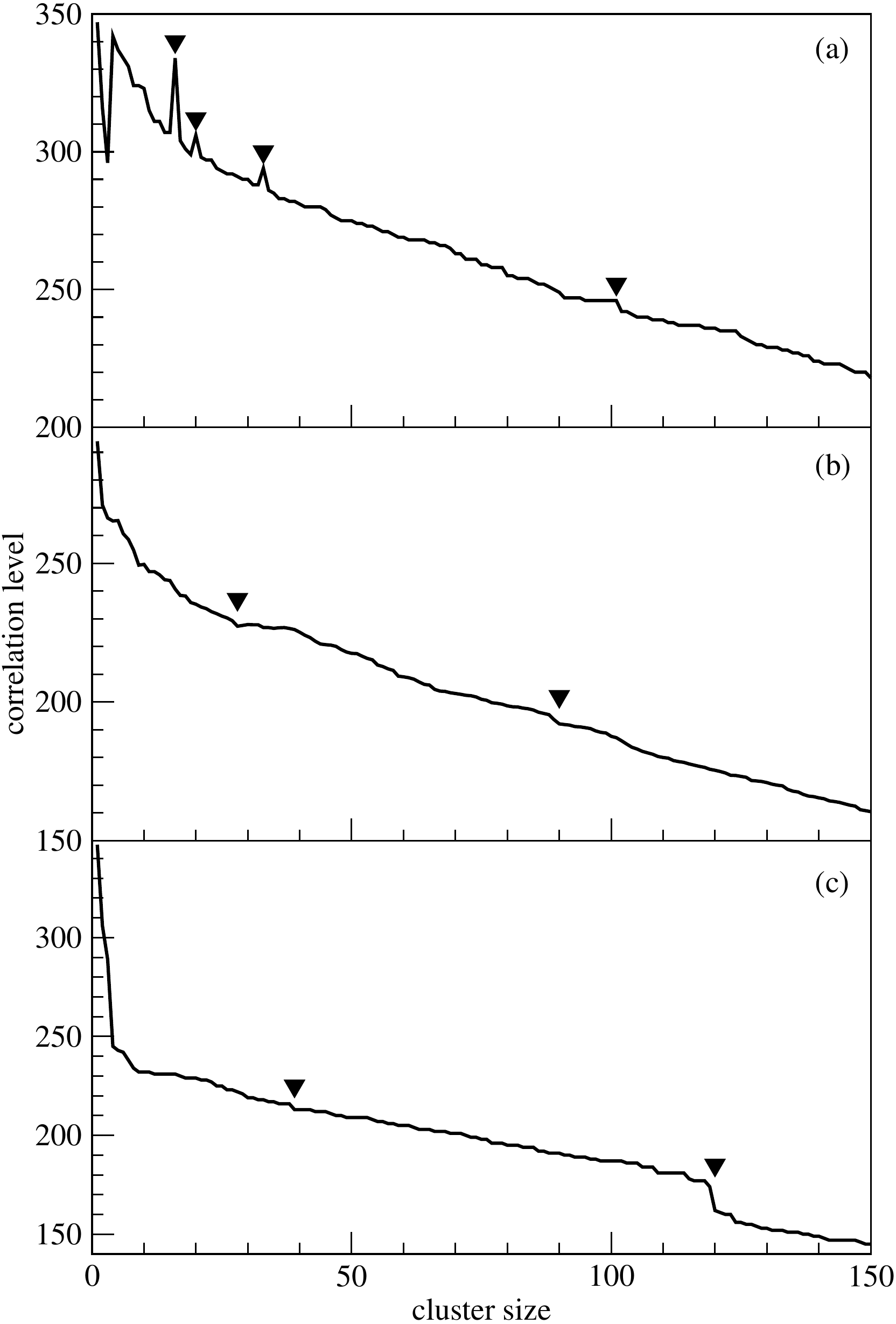}
\caption{Typical plots, taken from the SGX, of the correlation level
as a function of cluster size when we perform partial hierarchical
clustering using (a) the single link algorithm; (b) the average link
algorithm; and (c) the complete link algorithm.  Spikes in the
single-link partial hierarchical clustering history arise as the
growing cluster incorporates a small group of very strongly correlated
stocks (financial atoms), while sharp drops and kinks in all partial
hierarchical clustering histories arise when the growing cluster has
more or less exhausted stocks that are correlated with each other at a
certain correlation level (financial molecules), and starts
incorporating stocks that correlated with each other at a lower
correlation level.}
\label{fig:phctlinkages}
\end{figure}

To properly identify these financial atoms in a given market, we need
to be careful: just as core electronic densities are different for
different atoms, we suspect the correlation levels may also be
different for different financial atoms.  This suggests the need for
multiple thresholds, instead of a single threshold $c$, if we use
agglomerative hierarchical clustering to discover the financial atoms.
In order for the association between clusters in the hierarchical
clustering tree and atomic/molecular components in the financial
market to be statistically robust, we require the clusters to be
insensitive to variations in the individual thresholds.  To accomplish
this, one can imagine a sophisticated algorithm starting out with a
large number of thresholds, which are progressively merged as they are
slowly raised, based on how sensitive the clusters are to these
thresholds.  Such an algorithm, however, is difficult to implement,
and thus we adopt a simpler approach, where we start from a \emph{seed
cluster}, i.e. a core set of two very-strongly-correlated stocks, and
grow the hierarchical cluster outwards.  We call this procedure
\emph{partial hierarchical clustering}, and the sequence of
correlation levels at which new members are admitted into the growing
cluster its \emph{partial hierarchical clustering history}.  Figure
\ref{fig:phctlinkages} shows the typical partial hierarchical
clustering histories obtained using different linkage algorithms.
With all linkage algorithms, the statistical signature to look out for
is a sharp change in slope of the correlation level as a function of
cluster size.  This can take the form of sharp spikes, sharp drops, or
simply kinks.  These statistical signatures have different
interpretations for the different linkage algorithms.  We pick the
complete link algorithm for our analysis proper, because the sharp
drops and kinks in the correlation level can be most naturally
interpreted as cluster boundaries.  Compared to the statistical
signatures used by most existing statistical learning algorithms
\cite{Focardi2004QuantFin4p417, Gafichuk2004PhysicaA341p547,
Idicula2004SSRN634681, Aste2005ProcSPIE, Doherty2005ProcICSC,
Micciche2005ProcISSSP, Tumminello2005PNAS102p10421,
Allefeld2007PhysRevE76e066207, Rummel2007EuroPhysLett80e68004,
Tumminello2007EurPhysLett78e30006, Heimo2008PhysicaA387p5930,
Innocenti2008JPhysAMathTheor41e205101, Rummel2008PhysRevE77e016708,
Tumminello2008socph08094165, Ma2008SIAMRev50p413,
Alonson2006ComputStatistDataAnal51p762,
FruthwirthSchnatter2008JBusinessEconStats26p78,
Bauwens2007EconRev26p365, Corduas2008ComputStatistDataAnal52p1860,
Juarez2006CRiSM, Kullmann2000PhysicaA287p412}, we
believe ours is more intuitive.

\begin{figure}[htbp]
\centering
\includegraphics[scale=0.45,clip=true]{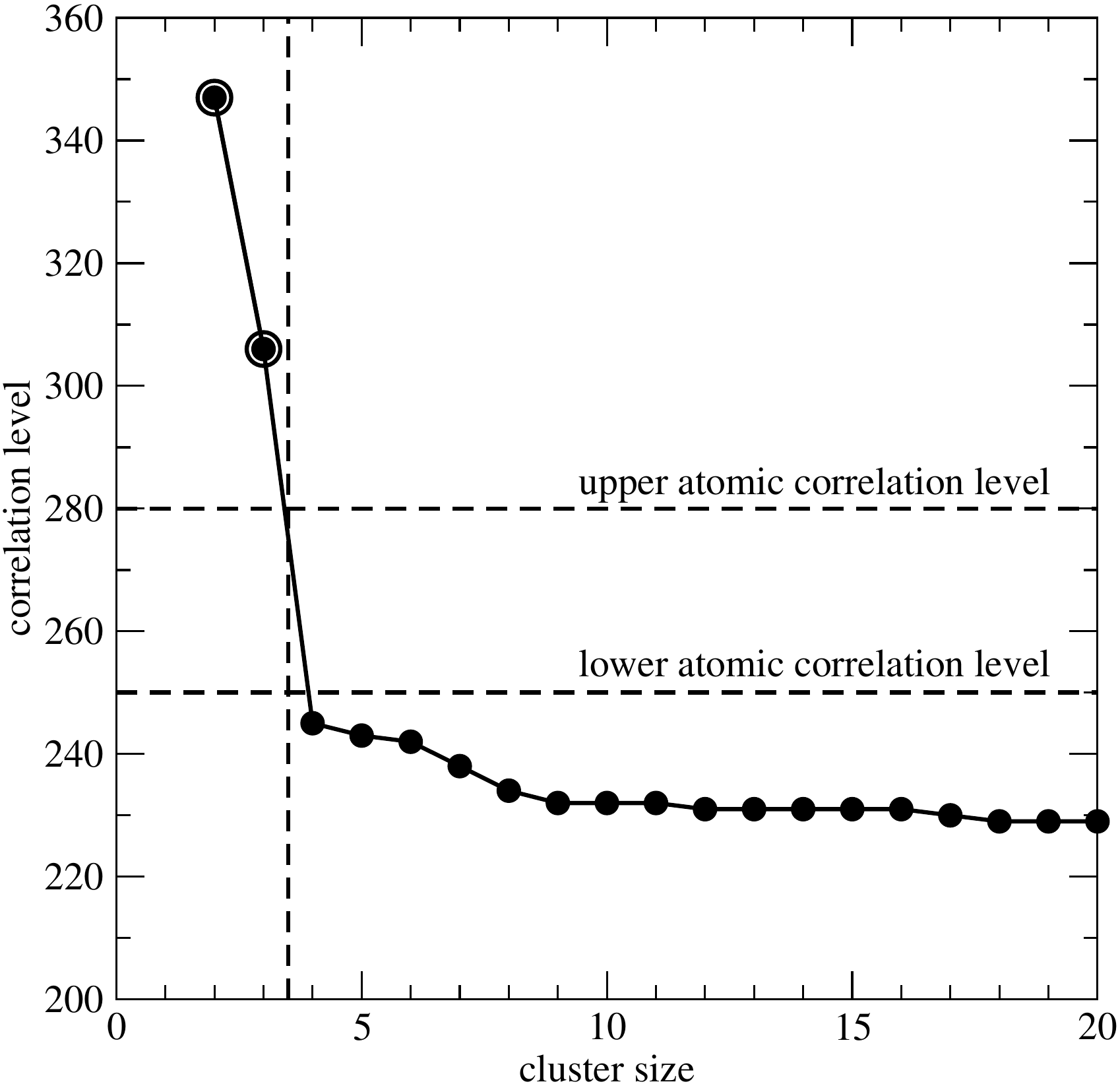}
\caption{A financial atom is identified from the complete-link partial
hierarchical clustering history of a given seed cluster by searching for
the sharpest change in slope at small cluster sizes.  Here, we show
the Singtel financial atom in the SGX, which consists of three stocks
(two of which are shown as decorated solid circles).}
\label{fig:phctatom}
\end{figure}

Since the total number of financial atoms in a given market is not
known, we discover them iteratively as follows.  First, let us use
$C^{(0)} = C$ to denote our long-time correlation matrix, and find the
maximum matrix element, $\max_{i,j} C^{(0)}(i, j) = C^{(0)}(i^{(1)},
j^{(1)})$.  Using $\{i^{(1)}, j^{(1)}\} = \{i_1^{(1)}, i_2^{(1)}\}$ as
our first seed cluster for complete-link partial hierarchical
clustering, we look for the sharpest change in the slope of the
correlation level as a function of cluster size, which marks the
`boundary' of the first candidate financial atom $\{i_1^{(1)},
i_2^{(1)}, \dots, i_{n(1)}^{(1)}\}$ (see Figure \ref{fig:phctatom}).
To find the second financial atom, we zero the $\{i_1^{(1)},
i_2^{(1)}, \dots, i_{n(1)}^{(1)}\}$ rows and columns of $C^{(0)}$, so
that the partial hierarchical clustering can proceed independently of
constituents of the first candidate financial atom.  We call this
modified correlation matrix $C^{(1)}$.  Again, we find $\max_{i,j}
C^{(1)}(i, j) = C^{(1)}(i^{(2)}, j^{(2)})$, and use $\{i^{(2)},
j^{(2)}\} = \{i_1^{(2)}, i_2^{(2)}\}$ as our second seed cluster for
complete-link partial hierarchical clustering.  From the sharpest
change in the slope of the correlation level, we identify the second
candidate financial atom $\{i_1^{(2)}, i_2^{(2)}, \dots,
i_{n(2)}^{(2)}\}$, and proceed to zero the $\{i_1^{(2)}, i_2^{(2)},
\dots, i_{n(2)}^{(2)}\}$ rows and columns of $C^{(1)}$ to get
$C^{(2)}$.  This iterative procedure mimics a full hierarchical
clustering having different thresholds, which we can use to find as
many candidate financial atoms as we like.  

For a candidate to qualify as a true financial atom, the constituent
atomic stocks must be more strongly correlated with each other than
they are with stocks outside of the financial atom.  This
`self-consistency' criterion defines an \emph{upper atomic correlation
level}, and if we are to adhere to this `self-consistency' criterion
strictly, only clusters of stocks with maximum long-time correlation
rising above the upper atomic correlation level can be considered to
have dynamically self-organized into \emph{strong} financial atoms.
This criterion is somewhat restrictive, so we looked for an
alternative criterion compatible with the self-organization observed.
Inspecting the various partial hierarchical clustering histories of
the most strongly-correlated seed clusters, we find the correlation
levels at which the \emph{first} non-atomic stock was admitted into
the growing clusters to be tightly clustered around a small number of
\emph{lower atomic correlation levels}.  We can therefore define
\emph{weak} financial atoms to be those candidate financial atoms
whose maximum correlations are greater than the average of the lower
atomic correlation levels.  In Table \ref{tab:numatoms}, we show the
upper atom correlation level, the average lower atomic correlation
level, and the number of financial atoms based on the two criterions,
in each of the five financial markets.  The compositions of strong
financial atoms found in the five equity markets are shown in Appendix
A.

\begin{table}[htbp]
\centering
\caption{The number of strong and weak financial atoms discovered in
the five equity markets.  Also shown are the upper atomic correlation
levels, the average lower atomic correlation levels, and the average
molecular correlation levels for the five equity markets, out of a
maximum possible value of 522 over the two-year observation period.
The maximum intra-cluster correlation of a cluster must exceed the
upper (lower) atomic correlation level for it to qualify as a strong
(weak) financial atom respectively.}
\label{tab:numatoms}
\vskip .5\baselineskip
\begin{tabular}{lccccc}
\hline
& NYSE & LSE & TSE & HKSE & SGX \\
\hline
Upper atomic correlation level & 420 & 370 & 380 & 340 & 280 \\
Number of strong financial atoms & 30 & 12 & 14 & 7 & 6 \\
Lower atomic correlation level & 379 & 355 & 348 & 306 & 262 \\
Number of financial atoms & 108 & 22 & 52 & 16 & 11 \\
Molecular correlation level & 349 & 310 & 320 & 269 & 211 \\
\hline
\end{tabular}
\end{table}

In all markets, we find small financial atoms comprising on the order
of 10 stocks.  In all markets except for the TSE, we find trivial
financial atoms consisting of different stocks issued by the same
company, e.g. VIA.A and VIA.B (both offered by Viacom) in the NYSE,
SDR and SDRt (both offered by Schroders plc) in the LSE, 0019 and 0087
(both offered by Swire Pacific) in the HKSE, and Singtel, Singtel 10,
and Singtel 100 (all offered by Singtel) in the SGX.  In the NYSE, LSE
and TSE, the constituent stocks of nearly all strong financial atoms
fall cleanly into the respective market sectors.  In the LSE, the
exception is financial atom 12, which consists of mostly insurance
stocks (Aviva, Prudential, Legal and General Group, Friends
Provident), but also contain one bank stock (Standard Chartered), and
one investment trust stock (Foreign \& Colonial Investment Trust).  In
the TSE, the exception is financial atom 9, which consists of real
estate stocks (Tokyo Tatemono Co, Urban, Creed Co, Pacific Holding
Inc), apart from Kenedix Inc, which is classified as belonging to the
Services market sector.  However, Kenedix Inc's four business segments
are all related to the real estate market.

In the HKSE and SGX, on the other hand, we find a significant
proportion of strong financial atoms having constituents that straddle
several market sectors (financial atoms 1 and 2 in the HKSE, and
financial atoms 4 and 6 in the SGX).  The emergence of these strong
cross-sector correlations can only be explained by the underlying
shareholder profiles, and confluential business interests.  For
example, in the HKSE financial atom 1, 40.23\% of the stake in Cheung
Kong is held by Hong Kong business tycoon Li Ka-Shing \cite{OSIRIS}.
Cheung Kong, in turn, owns 49.97\% of Hutchison Whampoa's shares.
Similarly, the Hong Kong and Shanghai Banking Corporation (HSBC) owns
62.14\% of the shares of the Hang Seng Bank, and holds a 42.01\% net
stake in Sun Hung Kai Properties.  These two pairs of stocks have
little in common in terms of ownership, and the main reason they are
so strongly-correlated is probably Cheung Kong and Hutchison Whampoa
buying over HSBC and Hang Seng's stakes in the e-commerce setup
iBusiness on July 27, 2007 \cite{CKHJul2007}.  The HKSE financial atom
1 also contains the MTR Corp, a statutory corporation owned by the
Hong Kong Government, and China Mobile, a state-owned enterprise of
the People's Republic of China.  The exact reasons for these two to be
strongly correlated with Cheung Kong-Hutchison Whampoa and Hang
Seng-Sun Hung Kai Properties are unclear (HSBC has a 0.15\% stake in
MTR Corp, but this is surely too little to account for the observed
correlations).  We believe these strong correlations may be related to
the perceived competition between China Mobile and Hutchison Whampoa,
which has a telecommunications arm, the announcement on April 13, 2006
that the MTR Corp signed an agreement with the Beijing municipal
government to build and operate a new Beijing rail line
\cite{APApr2006}, or the announcement on Nov 28, 2007 that Cheung Kong
won a HK\$7 billion MTR property project \cite{ReutersNov2007}.

Another interesting observation that we would like to highlight, is
the finding of more than one strong financial atom for some market
sectors in the LSE and TSE.  In contrast, we find at most one strong
financial atom per market sector in the NYSE.  In the LSE, we find a
total of four strong financial atoms belonging to the Investment
Trusts market sector, whereas in the TSE, we find two strong financial
atoms each in the Real Estate, Banks, and Non-Ferrous Metals market
sectors.  Looking at the example of the LSE financial atoms 4
(Scottish Investment Trust and Witan Investment Trust) and 5 (Monks
Investment Trust and Scottish Mortgage Investment Trust), we find that
Scottish Investment Trust and Witan Investment Trust have in common
ownership by the French holding company AXA, which holds 11.30\% and
16.00\% stakes in the two investment trusts respectively
\cite{OSIRIS}.  The shareholders they have in common also include the
Legal and General Gp (4.10\% and 4.59\% respectively) and Lloyds TSB
Gp (2.99\% and 0.26\% respectively).  In contrast, AXA is missing from
the lists of larger shareholders for Monks Investment Trust and
Scottish Mortgage Investment Trust, which feature instead Barclays
(5.99\% and 3.61\% respectively), as well as Legal and General Gp
(3.80\% and 3.60\% respectively).  For the example of the TSE
financial atoms 5 and 8, while the ownership makeups tell interesting
stories, we believe the main reason for their dynamics to be decoupled
is TSE financial atom 5 being made up entirely of large regional
banks, with operating revenue less than US\$1 billion, while TSE
financial atom 8 consists of even larger diversified banks, with
operating revenue significantly larger than US\$1 billion.

Finally, we note that strong foreign financial atoms are detected in
the NYSE and LSE, reflecting their more globalized character, but not
in the other three markets.   In the LSE, the only strong foreign
financial atom (atom 11) consists of two Korean stocks issued by
Samsung Electronics.  In the NYSE, on the other hand, there are a
total of 10 strong foreign financial atoms (the American Depository
Receipts (ADR) atoms).  In general, foreign financial atoms are
smaller compared to the typical local financial atoms.  Many of them
are also part of strong financial atoms in their home markets.  We
shall also see later, while LSE atom 11 is weakly correlated with the
rest of LSE, that the dynamics of some of the NYSE ADR atoms are
strongly coupled to that of many US atoms in the NYSE.

\subsection{Financial Molecules and Molecular Correlation Level}

\begin{figure}[htbp]
\centering
\includegraphics[scale=0.45,clip=true]{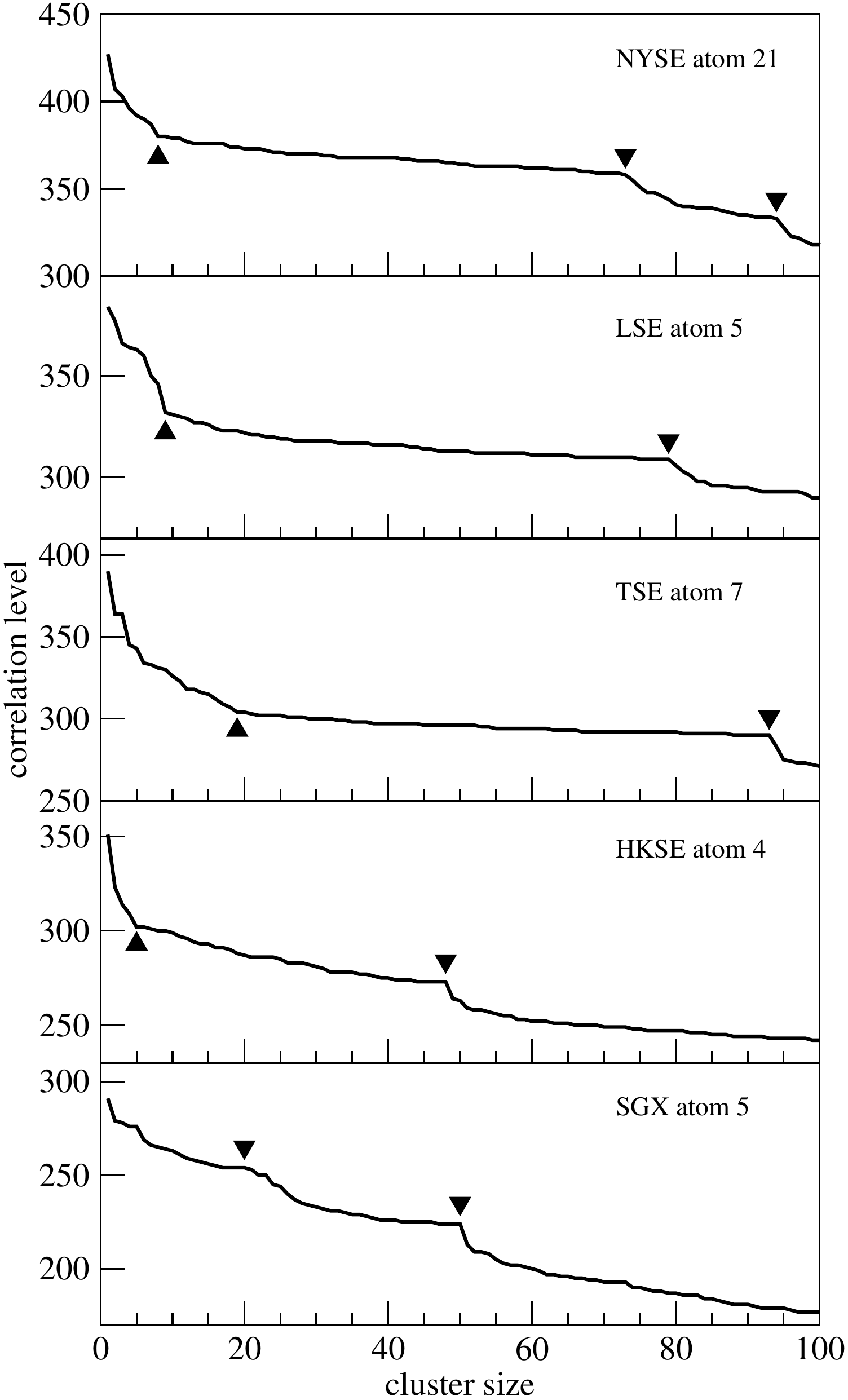}
\caption{Complete-link partial hierarchical clustering histories
starting from (top to bottom): financial atom 21 of the NYSE,
financial atom 5 of the LSE, financial atom 7 of the TSE, financial
atom 4 of the HKSE, and financial atom 5 of the SGX.  In these plots,
the most distinctive changes in slope of the correlation level, which
mark natural boundaries of financial molecules, are indicated by the
black triangles.}
\label{fig:phctmol}
\end{figure}

From Figure \ref{fig:phctlinkages}, we see statistical signatures
indicating the presence of higher level self-organization in real
financial markets.  We refer to self-organized structures one level
higher than financial atoms as \emph{financial molecules}.  Just as
molecules are made up of atoms, we expect financial molecules to also
be composed of financial atoms.  Again, we intuitively expect
financial atoms in the same financial molecule to be more strongly
correlated with each other, compared to financial atoms in different
financial molecules.  To find these entities, we perform complete-link
partial hierarchical clustering of the \emph{full} correlation matrix
--- so that all financial atoms will be considered --- starting from a
given financial atom.  We then examine the partial hierarchical
clustering history, and look out for sharp changes in the slope of the
correlation level as a function of cluster size (see Figure
\ref{fig:phctmol}).  These occur whenever a financial atom from a
different financial molecule is added to the growing cluster by the
partial hierarchical clustering procedure.  Going over the partial
hierarchical clustering histories of all strong financial atoms, 
we find the correlation levels at which the first non-molecular atom
was admitted into the growing clusters to be tightly clustered around
a small number of \emph{molecular correlation levels}.  To illustrate
the separation of atomic and molecular correlation levels that result
from the separation atomic and molecular time scales, we show the
average molecular correlation level in Table \ref{tab:numatoms}, for
easy comparison against the atomic correlation levels.

Using sharp changes in the slope of the correlation level-cluster size
plots as natural markers for molecular boundaries, we can determine
the collections of stocks making up various financial molecules in a
given market.  A molecular collection of stocks so identified
represents part or whole of a financial molecule.  In general, we need
to examine several such molecular fragments to deduce the complete
microscopic makeup of the financial molecule.  However, we find this
stock-level composition of the financial molecule less useful than its
atom-level composition.  If one or more stocks from a financial atom
is found in any molecular fragment of a financial molecule, we say
that the financial atom is a \emph{constituent atom} of the financial
molecule.  We call the list of constituent atoms, and their pattern of
correlations, the \emph{molecular structure} of a given financial
molecule.  Within this atomic caricature of the financial molecule,
the molecular fragments are groups of constituent atoms that are more
strongly correlated with each other than they are with other
constituent atoms.  As we shall illustrate below using the SGX
financial molecule as example, the molecular fragments that we
obtained starting from different constituent atoms, plus and minus
some non-atomic stocks, are consistent with the overall correlational
structure of the financial molecule.  This tells us that financial
molecules are robust self-organized features in the dynamics of real
financial markets.

Within a financial molecule, all constituent atoms are correlated with
each other, but more strongly with some, and less strongly with
others.  To produce an intuitive picture of the correlational
structure within the financial molecule, we draw bonds between
financial atoms that are strongly correlated.  There is no unique way
to do this, but we find the following rules producing
rather informative diagrams:
\begin{enumerate}

\item area of a financial atom is proportional to the number of stocks
it contains; 

\item solid circle for strong financial atoms, and dashed circle for
weak financial atoms;

\item select a upper correlation level $c_1$ and lower correlation
level $c_2$ based on the intra-molecular correlations \cite{C1C2},
and draw:

	\begin{enumerate}

	\item a thick bond between two constituent atoms if their
	maximum and minimum interatomic correlation exceeds $c_1$ and $c_2$
	respectively;

	\item a thin bond between two constituent atoms if their maximum
	interatomic correlation exceeds $c_1$, but their minimum
	interatomic correlation is below $c_2$;

	\item a dashed bond between two constituent atoms if their maximum
	interatomic correlation is between $c_1$ and $c_2$;

	\item no bond between two constituent atoms if their maximum
	interatomic correlation is below $c_2$;

	\end{enumerate}

\item if a constituent atom is bonded to other constituent atoms by
solid as well as dashed bonds, remove the dashed bonds;

\item shade a constituent atom light gray, if its correlations with
non-atomic components exceed $c_2$.  

\end{enumerate}

\subsubsection{Financial Molecules in the SGX and HKSE}

\begin{figure}[htpb]
\centering
\includegraphics[scale=0.4,clip=true]{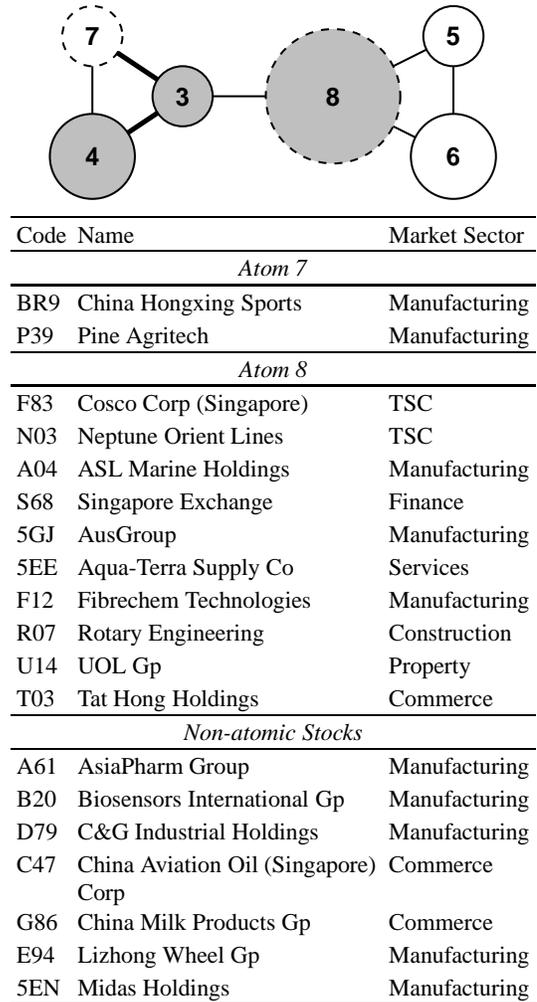}
\vskip .5\baselineskip
\begin{tabular}{lll}
\hline
Code & Name & Market Sector \\
\hline
\multicolumn{3}{c}{\emph{Atom 7}} \\
\hline
BR9 & China Hongxing Sports & Manufacturing \\
P39 & Pine Agritech & Manufacturing \\
\hline
\multicolumn{3}{c}{\emph{Atom 8}} \\
\hline
F83 & Cosco Corp (Singapore) & TSC \\
N03 & Neptune Orient Lines & TSC \\
A04 & ASL Marine Holdings & Manufacturing \\
S68 & Singapore Exchange & Finance \\
5GJ & AusGroup & Manufacturing \\
5EE & Aqua-Terra Supply Co & Services \\
F12 & Fibrechem Technologies & Manufacturing \\
R07 & Rotary Engineering & Construction \\
U14 & UOL Gp & Property \\
T03 & Tat Hong Holdings & Commerce \\
\hline
\multicolumn{3}{c}{\emph{Non-atomic Stocks}} \\
\hline
A61 & AsiaPharm Group & Manufacturing \\
B20 & \parbox[t]{4cm}{\raggedright Biosensors International Gp} &
Manufacturing \\
D79 & \parbox[t]{4cm}{\raggedright C\&G Industrial Holdings} & 
Manufacturing \\
C47 & \parbox[t]{4cm}{\raggedright China Aviation Oil (Singapore)
Corp} & Commerce \\
G86 & China Milk Products Gp & Commerce \\
E94 & Lizhong Wheel Gp & Manufacturing \\
5EN & Midas Holdings & Manufacturing \\
\hline
\end{tabular}
\caption{The six-atom SGX financial molecule comprising the strong
financial atoms 3, 4, 5 and 6, as well as the weak financial atoms 7
and 8 (see table above for compositions).  This is deduced from the first
natural boundaries in the partial hierarchical clustering histories of
the strong financial atoms 3, 4, 5, and 6.  The bonds are drawn with $c_1
= 280$ and $c_2 = 262$.  In the table above, TSC is an acronym for the
transportation, storage, and communications market sector.}
\label{fig:SGXcmol3}
\end{figure}

In the SGX, we find only one financial molecule, whose composition
depends on the level of statistical significance we desire.  When we
look at the partial hierarchical clustering histories starting from
the strong financial atoms 3, 4, or 5, we find two natural molecular
boundaries (see bottommost panel in Figure \ref{fig:phctmol}).
Depending on the starting financial atom, the statistically less
significant first boundary suggests different lists of constituent
atoms: (3, 4, 5, 6, 7, 8) when we start from strong financial atoms 3
and 5, and (3, 4, 7, 8) when we start from strong financial atom 4.
From the partial hierarchical clustering history of strong financial
atom 6, we identify only one natural molecular boundary, and list of
constituent atoms suggested by this sole molecular boundary is (5, 6).
Based on these observations, we conclude that the partial
hierarchical clustering procedure finds different parts of the
financial molecule if we start from different constituent atoms.
These different parts, however, are consistent with the molecular
structure of the six-atom financial molecule shown in Figure
\ref{fig:SGXcmol3}, which we deduce by taking the union of the
different lists of constituent atoms, and drawing bonds based on the
rules listed above.  Different starting constituent atoms also give us
different lists of constituent non-atomic stocks.  Again, we take the
union of these lists, and find that the constituent non-atomic stocks
are most strongly correlated with the financial atoms 3, 4, and 8.
Because their strong sub-atomic correlations with multiple financial
atoms, we can interpret these constituent non-atomic stocks as
`bonding' stocks.

\begin{figure}[htpb]
\centering
\includegraphics[scale=0.4,clip=true]{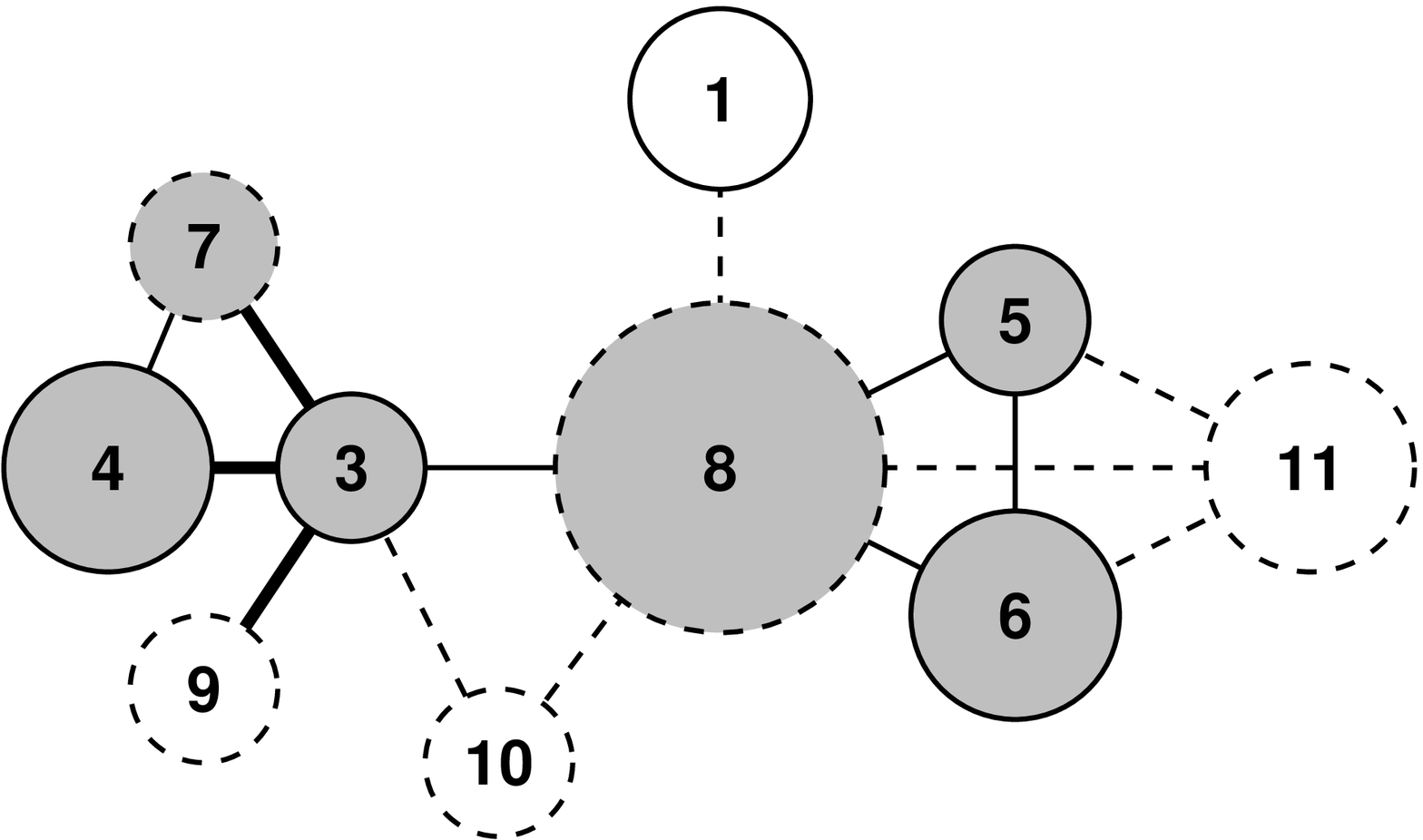}
\vskip .5\baselineskip
\begin{tabular}{lll}
\hline
Code & Name & Market Sector \\
\hline
\multicolumn{3}{c}{\emph{Atom 9}} \\
\hline
L28 & Longcheer Holdings & Services \\
L46 & \parbox[t]{4cm}{Luzhou Bio-chem Technology} & Manufacturing \\
\hline
\multicolumn{3}{c}{\emph{Atom 10}} \\
\hline
F20 & Federal International (2000) & Commerce \\
T06 & Tiong Woon Corp Holdings & TSC \\
\hline
\multicolumn{3}{c}{\emph{Atom 11}} \\
\hline
A17U & \parbox[t]{4cm}{\raggedright Ascendas Real Estate Investment
Trust} & Property \\
C61U & CapitaCommercial Trust & Property \\
C38U & \parbox[t]{4cm}{\raggedright CapitaMall Trust Management} & 
Property \\
P27 & Parkway Holdings & Services \\
\hline
\multicolumn{3}{c}{\emph{Additional Non-atomic Stocks}} \\
\hline
CEGL & China Essence Gp & Manufacturing \\
E13 & Ellipsiz & Services \\
5DN & Ezra Holdings & TSC \\
H80 & Hongwei Technologies & Manufacturing \\
O10 & Orchard Parade Holdings & Property \\
P11 & Pacific Andes (Holdings) & Manufacturing \\
S63 & \parbox[t]{4cm}{\raggedright Singapore Technologies Engineering} & 
Manufacturing \\
Y34 & Sunshine Holdings & Property \\
5FG & Swissco International & TSC \\
\hline
\end{tabular}
\caption{The 10-atom SGX financial molecule, within which the six-atom
financial molecule shown in Figure \ref{fig:SGXcmol3} is nested,
deduced from the second natural boundaries in the partial hierarchical
clustering histories of the strong financial atoms 3, 4, and 5.
Compositions of the three additional participating weak financial
atoms are shown in the table above, as are additional non-atomic
stocks.  The bonds are drawn with $c_1 = 280$ and $c_2 = 262$.}
\label{fig:SGXcmol3a}
\end{figure}

In Figure \ref{fig:SGXcmol3}, we see that this six-atom financial
molecule consists of two three-atom clusters, (3, 4, 7) and (5, 6, 8),
connected by a single bond between atoms 3 and 8.  Inspection of
atomic compositions tells us that the property atom 5, banking atom 6,
and shipping atom 8 consist mostly of local companies, whereas the
manufacturing atoms 3, 4 and 7 consist only of Chinese companies
listed on the SGX or China-related local companies.  Most of the
non-atomic stocks are also stocks of Chinese or China-related
companies.  The larger 10-atom financial molecule shown in Figure
\ref{fig:SGXcmol3a}, suggested by the statistically more significant
second boundaries in the partial hierarchical clustering histories of
financial atoms 3, 4, and 5, tells an even more intricate story.
Apart from the nested six-atom molecular core shown in Figure
\ref{fig:SGXcmol3}, we find also the participation of financial atoms
1, 9, 10, and 11.  In this larger financial molecule, we find the same
basic topology: a cluster of China related atoms $\{3, 4, 7, 9\}$, and
a cluster of local atoms $\{1, 5, 6, 8, 11\}$.  Apart from the direct
bonding between financial atoms 3 and 8, the two clusters are also
bonded indirectly through the weak bonds between atoms 3 and 8 with
the TSC atom 10.  We believe it is likely that in 2005 or 2006, the
two clusters might actually represent two distinct financial
molecules, which became increasingly correlated with each other in the
period leading up to, and beyond, the end-Feb 2007 market crash known
as the \emph{Chinese Correction}.

\begin{figure}[htpb]
\centering
\includegraphics[scale=0.4,clip=true]{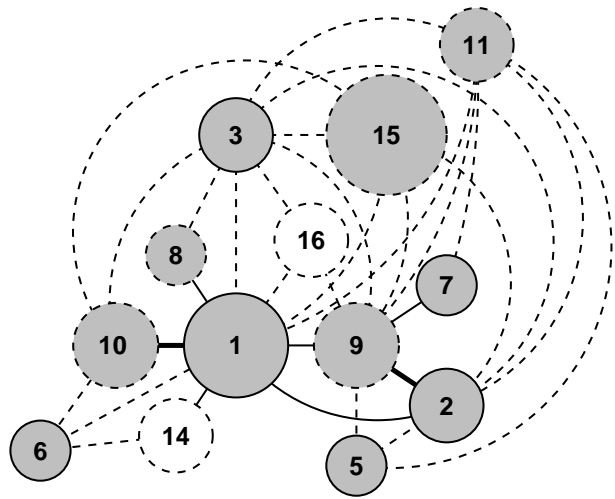}
\caption{The 13-atom molecule in the HKSE, drawn with $c_1 = 340$ and
$c_2 = 306$.  This molecule consists of two clusters of atoms, $\{1,
8, 10, 14, (6)\}$, and $\{2, 7, 9, (5), (11)\}$, connected by direct
bonding, as well as indirect bonding through atoms 3, 15 and 16.  The
first cluster consists almost exclusively of local atoms, while the
second cluster are exclusively China atoms.  Atoms 3 and 16 are both
local atoms, but their constituent companies have very heavy property
investments in China, while atom 15 is partially-local and partially
Chinese.}
\label{fig:HKSEcmol2}
\end{figure}

In the HKSE, we find also a single 13-atom financial molecule shown in
Figure \ref{fig:HKSEcmol2}.  Its molecular structure is considerably
more complex than the SGX financial molecule, but we can still make
out two molecular cores, $\{1, 6, 8, 10, 14\}$ and $\{2, 5, 7, 9\}$,
as well as a group $\{3, 11, 15, 16\}$ of bridging atoms.  Inspection
of the atomic compositions within the first molecular core, we
realized that $\{1, 6, 8, 10, 14\}$ are all local atoms, whose
constituent stocks are issued by companies based in Hong Kong.  Apart
from financial atom 14, which is a banking and finance atom, the rest
are all property atoms.  The second molecular core $\{2, 5, 7, 9\}$,
on the other hand, contains only Chinese atoms, whose constituent
stocks are issued by companies based in China.  Unlike the local
molecular core, atoms from the Chinese molecular core are from a
variety of industries, ranging from banking and finance (2, 9), to oil
and energy (7), to mining and metals (5).  In the bridging group of
financial atoms, we find a mix between local and Chinese atoms,
primarily from the property market (3, 15, 16) and mining industry
(11).  In addition to indirect bonding of the two molecular cores
through the bridging group of atoms, we also find strong direct bonds
between the local atom 1 and Chinese atoms 2 and 9.  The non-atomic
stocks are also of mixed local and Chinese origins, representing a
mixture of industries.  These are strongly correlated with nearly
every constituent atom, and can most appropriately be interpreted as a
`valence cloud' of the financial molecule.  Unlike the situation in
the SGX, many Hong Kong companies have direct business involvements in
China, so it is perhaps not surprising to find strong correlations
between the local and Chinese molecular cores.  However, we believe
that prior to 2006, atoms within the two molecular cores probably
constitute several distinct financial molecules, whose movements
become increasingly intertwined over 2006, culminating in the Chinese
Correction of Feb 2007.  We also believe that the industry cross
section reflected in the HKSE financial molecule will provide
important clues to understanding the Chinese Correction.

\begin{figure}[htpb]
\centering
\includegraphics[scale=0.4,clip=true]{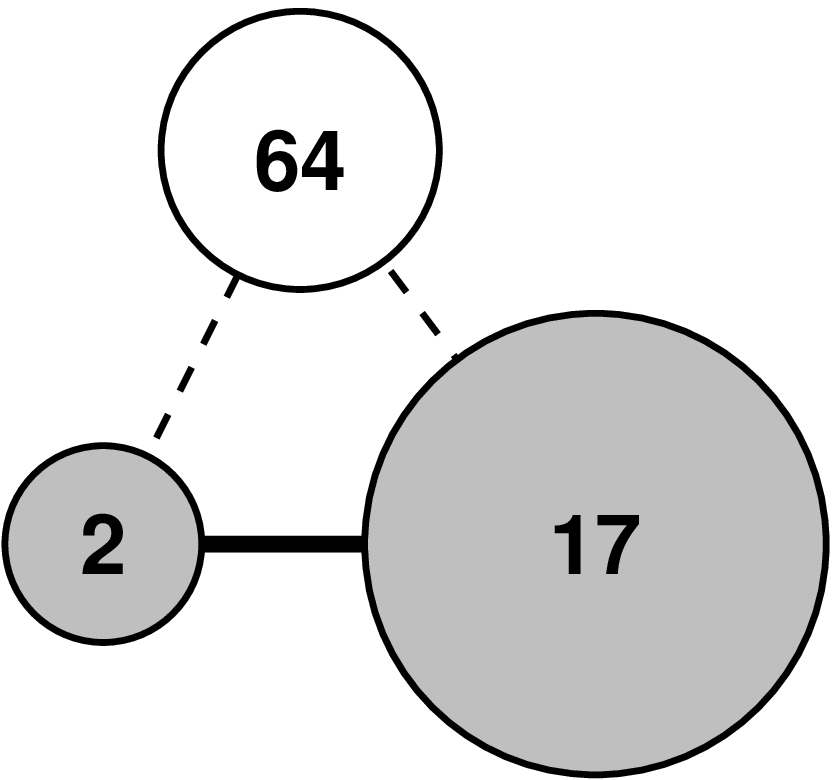}
\vskip .5\baselineskip
\begin{tabular}{lll}
\hline
Code & Name & Market Sector \\
\hline
\multicolumn{3}{c}{\emph{Atom 64}} \\
\hline
GS & Goldman Sachs Gp Inc & Financial \\
LBI & Lehman Brothers Inc & Financial \\
MER & Merrill Lynch \& Co Inc & Financial \\
RJF & Raymond James Financial Inc & Financial \\
\hline
\multicolumn{3}{c}{\emph{Non-Atomic Stocks}} \\
\hline
MHO & M/I Homes Inc & Homebuilding \\
NVR & NVR Inc & Homebuilding \\
MHK & Mohawk Industries Inc & Home Furnishing \\
DY & Dycom Industries Inc & Heavy Construction \\
\hline
\end{tabular}
\caption{The 3-atom molecule in the NYSE, which consists of very
strongly-bonded homebuilding atoms 2 and 17, both of which are weakly
bonded to the investment atom 64.  Four non-atomic stocks are
correlated with atoms 2 and 17 strongly above market level
correlations.  Two of these, M/I Homes Inc (MHO) and NVR Inc (NVR),
are homebuilders, while Mohawk Industries Inc (MHK) is in the home
furnishing business.  The last non-atomic stock is Dycom Industries
Inc (DY), who is a provider of specialty engineering and construction
contracting services.  Because of its unique correlational signatures,
the bonds in this molecule are not drawn using the rules described
above, but through visual inspection of the actual correlation
submatrix elements.}
\label{fig:NYSEcmol2}
\end{figure}

\begin{figure}[htpb]
\centering
\includegraphics[scale=0.4,clip=true]{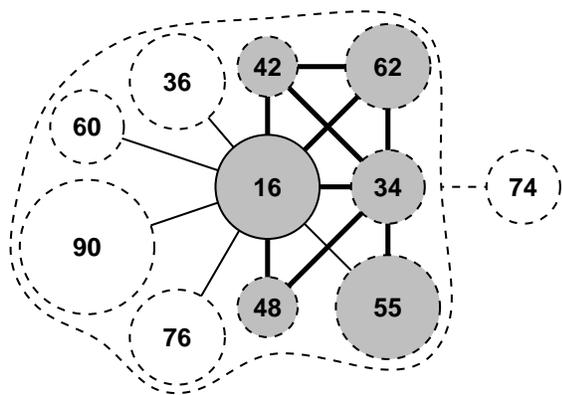}
\caption{The 11-atom real estate financial molecule in the NYSE, drawn
with $c_1 = 403$ and $c_2 = 377$.  This financial molecule consists of
a core of six financial atoms (6, 34, 42, 48, 55, and 62, shaded
gray), surrounded by a inner shell of four financial atoms (36, 60,
76, 90) bonded to financial atom 16, and an outer shell consisting of
financial atom 74.  The six core real estate investment trust (REIT)
atoms are in the more generic office, retail, residential and
industrial realty businesses, whereas the four shell REIT atoms are
all rather specialized, either in the realty needs of the healthcare,
hospitality or retail mall sectors.  Atom 36, which is bonded only to
REIT atom 16, is a mutual funds atom consisting largely of the Cohen
\& Steers group of realty related funds.  In the above diagram, the
dashed bond to the dashed envelope is a shorthand indicating that atom
74 is weakly bonded to every atom within the envelope.}
\label{fig:NYSEcmol16} 
\end{figure}

\subsubsection{Financial Molecules in the NYSE, LSE and TSE}

For the larger markets, we find more financial molecules, with a
variety of sizes and compositions.  One of the smallest financial
molecule we find in the NYSE consists of the atoms $\{2, 17, 64\}$
(shown in Figure \ref{fig:NYSEcmol2}).  This financial molecule is
interesting because the strong financial atoms 2 and 17 are both in
the homebuilding market sector, whereas component stocks in the weak
financial atom 64 are in the Investment Services and Investment Trusts
industries.  Seeing that most component stocks in atom 64 got into
trouble in 2008, this correlational structure is perhaps a tell-tale
sign of the Subprime Crisis.  Looking for further signatures of the
Subprime Crisis, we found the large 11-atom financial molecule shown
in Figure \ref{fig:NYSEcmol16}.  This financial molecule contains
almost exclusively real estate and property related atoms, except for
atom 36, which is a mutual funds atom consisting mostly of Cohen \&
Steers funds.  Based on the inter-atomic correlations, as well as
correlations between the constituent atoms and constituent non-atomic
stocks, we find in this financial molecule a molecular core consisting
of the office, retail, residential and industrial real estate
investment trust (REIT) atoms 16, 34, 42, 48, 55, and 62.  This
strongly-correlated molecular core is surrounded by a inner shell of
four atoms (36, 60, 76, 90), bonded directly to atom 16, and an outer
shell consisting of atom 74, which is weakly bonded to every atom in
the financial molecule.  We see here that the mutual funds atom 36 is
strongly correlated with the molecular core through atom 16, but not
with the remaining four shell REIT atoms.  Inspecting their atomic
compositions, we also see that the four shell REIT atoms are all
rather specialized, either in the realty needs of the healthcare,
hospitality or outlet mall sectors.

\begin{figure}[htpb]
\centering
\includegraphics[scale=0.4,clip=true]{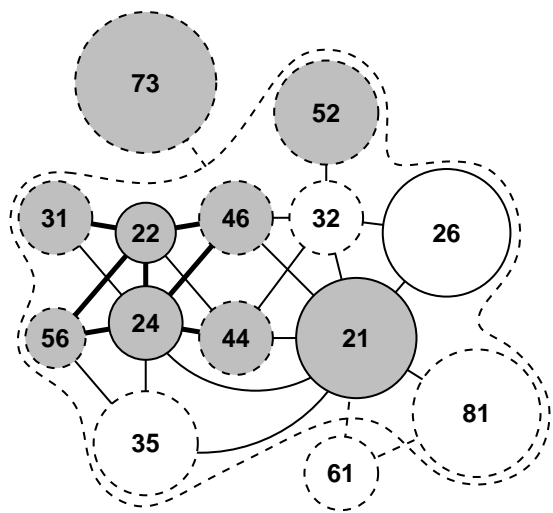}
\caption{The 14-atom US oil \& gas financial molecule in the NYSE,
drawn with $c_1 = 409$ and $c_2 = 383$.  This financial molecule
appears to consist of two cores: a very tight drilling \&
exploration/equipment \& services core consisting of atoms 22, 24, 31,
44, 46, 56, and a looser oil \& gas core consisting of atoms 21, 26,
32, and 81.  As shown using the shorthand introduced in Figure
\ref{fig:NYSEcmol16}, the equipment \& services atom 73 is weakly
bonded to every atom in the dashed envelope.  Surprisingly, we find
the participation of the coal atom 61 in this oil \& gas molecule.} 
\label{fig:NYSEUSoil}
\end{figure}

\begin{figure}[htpb]
\centering
\includegraphics[scale=0.4,clip=true]{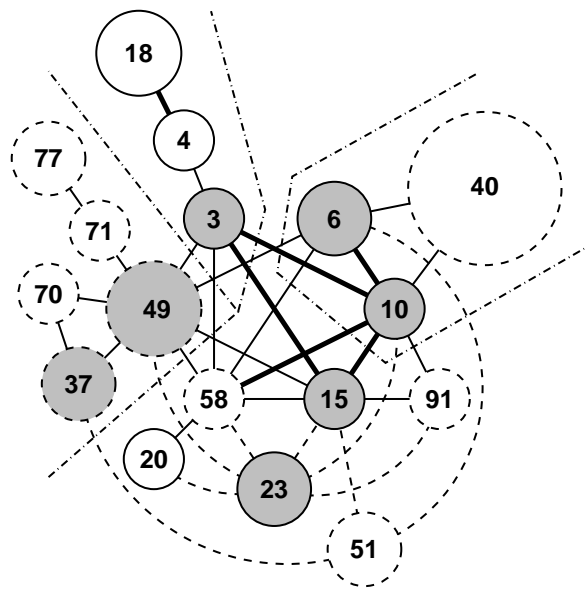}
\caption{The 17-atom ADR financial molecule in the NYSE, drawn
with $c_1 = 380$ and $c_2 = 370$.  The dashed-dotted lines demarcate
the distinct molecular cores: foreign oil (3, 4, 18), mining (6, 10,
40), and emerging market funds and banks (37, 49, 70, 71, 77).  The
remainder of the molecule comes from two groups of atoms: European
banks (15, 51), and an assortment of Brazilian companies (20, 23, 58,
91).  This multi-sector molecule is exceptional in the NYSE, where we
find financial molecules coming from one, or a few closely related
market sectors.}
\label{fig:NYSEADR}
\end{figure}

Another large financial molecule we found in the NYSE is the 14-atom
oil and gas molecule shown in Figure \ref{fig:NYSEUSoil}.  All atoms
in this molecule contain local stocks, and all except one are in the
closely related oil and gas (21, 26, 32, 81), drilling and exploration
(22, 31, 46, 73), equipment and services (24, 44, 52, 56, 73),
refining and marketing (35, 81) industries.  The exception is atom 61,
which is a coal atom.  The topology of this molecule suggests a
drilling \& exploration and equipment \& services molecular core, with
very strong internal correlations, coupled strongly to a oil \& gas
molecular core, with strong internal correlations.  Carefully
analyzing the correlations this local oil \& gas molecule have with
other financial atoms in the NYSE, we find that it is strongly
correlated with the 17-atom ADR molecule shown in Figure
\ref{fig:NYSEADR}.  This ADR molecule consists of three distinct
molecular cores.  The first is a foreign oil \& gas molecular core
comprising the ADR atoms 3, 4, and 18.  Of these three, the European
oil \& gas atoms 4 and 18 are the most strongly correlated with each
other.  The second is a mining molecular core, which consists of the
foreign mining atoms 6 and 10, and the US mining atom 40.  The third
molecular core consists of atoms associated with funds for and banks
in emerging markets.  In particular, the linear subcore (49, 71, 77)
are all fund atoms, while the cyclic subcore consists of Indian (37)
and Korean (70) bank atoms.  It is through the foreign oil \& gas
molecular core, and also partially through the mining molecular core,
that we find the strongest correlations between the ADR molecule and
the US oil \& gas molecule.  The remainder of the ADR molecule has a
rather complex composition, with 15 and 51 being European bank atoms,
and 20 (beer), 23 (telecommunications), 58 (steel), 91 (energy) being
Brazilian atoms.  It is plausible that in the past, these five
subclusters might represent five distinct financial molecules that
somehow became increasingly correlated in the period leading up to the
global financial crisis.

\begin{figure}[htpb]
\centering
\includegraphics[scale=0.4,clip=true]{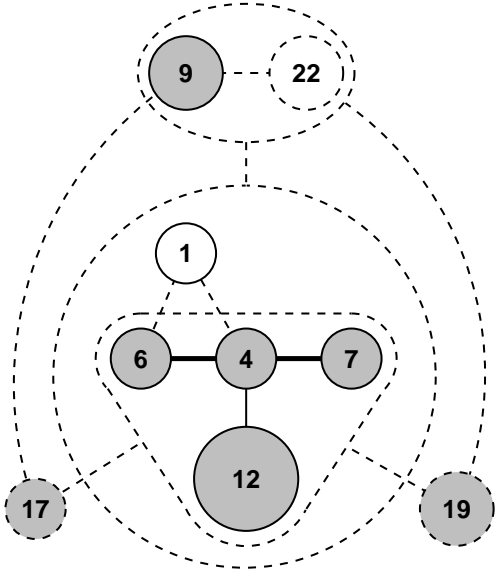}
\caption{The 9-atom investment trusts financial molecule in the LSE,
drawn with $c_1 = 376$ and $c_2 = 348$.  This molecule consists of an
inner core of investment trusts (1, 4, 6, 7) and insurance (12) atoms,
and an outer shell of investment trust atoms (9, 17, 19).
Surprisingly, the aerospace \& defense atom 22 is included as a member
of this molecule.}
\label{fig:LSEmol4}
\end{figure}

\begin{figure}[htpb]
\centering
\includegraphics[scale=0.4,clip=true]{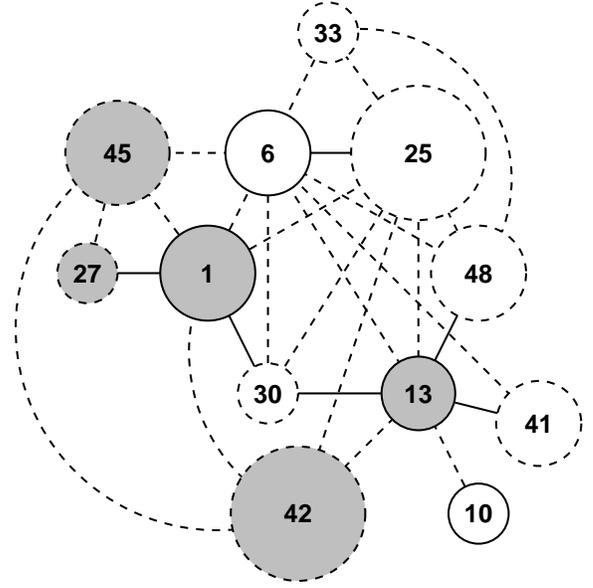}
\caption{The 12-atom multi-sector financial molecule in the TSE, drawn
with $c_1 = 370$ and $c_2 = 350$.  This molecule consists of two
molecular cores, (6, 25, 33), and (1, 10, 13, 27, 30, 41, 48), as well
as a group of bridging atoms (42, 45).  The industries represented in
this molecule are: steel, iron and metal mining (6, 10, 13, 25, 33,
41, 48), wholesale trade and distributing (1, 30), and heavy
machineries (41, 42, 45).  Surprisingly, we find that: (i) the (10,
13, 41, 48) of metal mining atoms are more strongly correlated with
the trading and distributing atom 30, than they are with the main
steel and iron group of atoms (6, 25, 33), inspite of the fact that
atom 48 consists mainly of steel production stocks; (ii) the heavy
machineries atom 27 is more strongly correlated with the wholesale
trade atom 1 than it is with the other heavy machineries atoms 41, 42,
and 45; (iii) the non-atomic stocks are mostly from the steel and
heavy machineries sectors, but have stronger correlations with the
wholesale trade atom 1, the non-ferrous metal atom 33, and the heavy
machineries atom 27, 42, and 45, and not with the steel and iron atoms
6, 25, and 33.}
\label{fig:TSEmol1}
\end{figure}

In the LSE and TSE, the financial molecules found are generally
smaller than those found in the NYSE, and closer in size to the
financial molecules found in the HKSE and SGX.  For example,
the LSE investment trusts financial molecule shown in Figure
\ref{fig:LSEmol4} comprises nine financial atoms, while the TSE heavy
industry molecule shown in Figure \ref{fig:TSEmol1} comprises 12
financial atoms.  The structure of the LSE investment trusts molecule
is relatively simple, with an inner core consisting of the investment
trusts (1, 4, 6, 7) and insurance (12) atoms, and an outer shell
consisting of the investment trusts atoms 9, 17, and 19.  All
non-atomic stocks are from the investment trusts and insurance
industries, and the only surprising member of this molecule is the
aerospace and defense atom 22.  The TSE heavy industry molecule, on
the other hand, has a significantly more complex structure, consisting
apparently of two molecular cores, (6, 25, 33), and (1, 10, 13, 27,
30, 41, 48) as well as a group of bridging atoms (42, 45).  Closer
inspection, however, reveal a mismatch between the bonding structure
and the industry distribution.  In this TSE financial molecule, we
find three major market sectors: steel, iron and metal mining (6, 10,
13, 25, 33, 41, 48), wholesale trade and distributing (1, 30), and
heavy machineries (41, 42, 45).  It is interesting that the (10, 13,
41, 48) group of metal mining atoms are more strongly correlated with
the trading and distributing atom 30, than they are with the main
steel and iron group of atoms (6, 25, 33).  This is inspite of the
fact that atom 48 consists of mostly steel production stocks.  It is
also surprising that the heavy machineries atom 27 is more strongly
correlated with the wholesale trade atom 1 than it is with the other
heavy machineries atoms, 41, 42, and 45.  Based on the industry makeup
of this financial molecule, it is not surprising to find that its
non-atomic stocks are mostly from the steel and heavy machineries
market sectors.  However, we find it surprising that the non-atomic
steel stocks are not strongly correlated with the steel and iron atoms
6, 25, and 33, but with the wholesale trade atom 1, the non-ferrous
metal atom 13, and the heavy machineries atoms 27, 42, and 45.

While the molecular structures in the LSE and TSE hold some surprises,
we find none of the extraordinary correlational structures seen in the
NYSE, HKSE, and SGX, which we believe offer insights into such
large-scale market events as the Chinese Correction and the Subprime
Crisis.  We believe that as a general rule, statistical signatures
preceding a market crash can only be seen in the market(s) of origin.
There should be little, or no symptoms at all, of such market crashes
in the other markets that follow the market of origin into decline.
However, because of strong coupling between global financial markets,
we believe it might be possible to see statistical signatures of
`aftershocks' that play out simultaneously in all markets after the
initial market crash.

\subsection{Financial Supermolecules}

\begin{figure}[htpb]
\centering
\includegraphics[scale=0.45,clip=true]{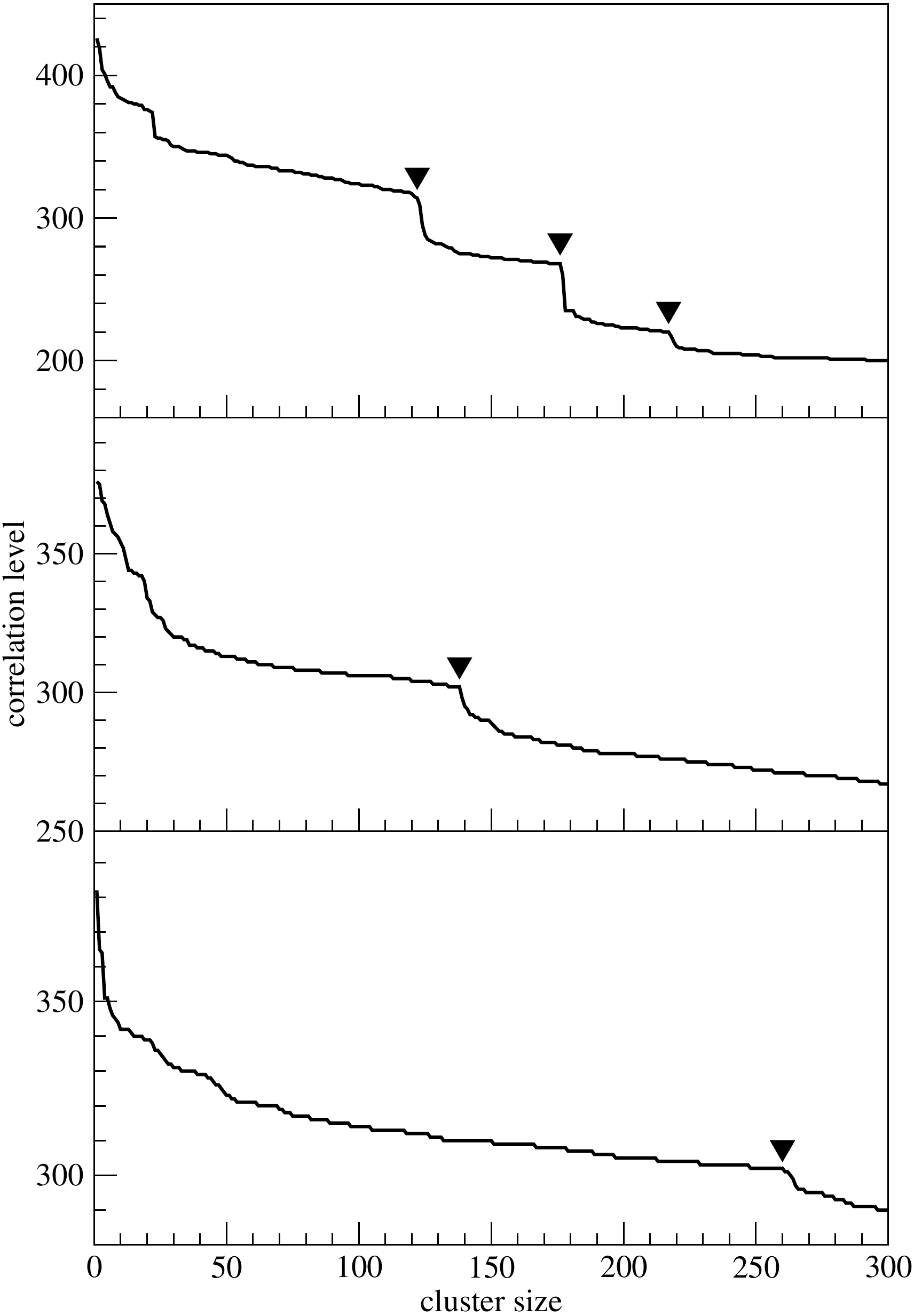}
\caption{Complete-link partial hierarchical clustering histories
showing statistical signatures of self-organization at the financial
supermolecule level in large mature markets, starting from (top to
bottom): atom 22 of the NYSE, atom 7 of the LSE, and atom 13 of the
TSE.  The natural boundaries of the financial supermolecules are
indicated by the black triangles.}
\label{fig:supermolecule}
\end{figure}

In all markets, we find further self-organization at the level of
hundreds of stocks.  In markets where we find more than one financial
molecule (NYSE, LSE, and TSE), these can be interpreted as bound
states of financial molecules, and thus we call them \emph{financial
supermolecules}.  From a finance point of view, the different market
sectors in a developed economy are subjected to differing degrees of
regulation, and as such, it is perhaps not surprising to find all
stocks within the same market sector to be more correlated with each
other, than with stocks from another market sector.  This naturally
suggests that the dynamics of an entire market sector self-organizes
into a financial supermolecule \cite{Coronello2005ActaPhysPolB36p2653,
Coronello2007ProcSPIE66010T, Kim2005PhysRevE72e046133,
Onnela2003PhysRevE68e056110}.  From the top
panel in Figure \ref{fig:supermolecule}, we find three natural
boundaries for a financial supermolecule in the NYSE, with the second
boundary at cluster size of 176 stocks being the most significant
statistically.  If we analyze the composition of this financial
supermolecule, starting from any of the natural boundaries, we find
that this supermolecule is primarily the bound state between the US
oil \& gas molecule (shown in Figure \ref{fig:NYSEUSoil}) and the ADR
molecule (shown in Figure \ref{fig:NYSEADR}).  If we use the third
boundary as our definition of the supermolecule, then we find it
including other US atoms from the oil \& gas (67, 105), steel (50,
95), emerging market funds (41, 45), industrial machinery (102),
aerospace and defense (89), and construction materials (106) sectors.
Most of these peripheral atoms are closely related to the US oil \&
gas molecule, or to the ADR molecule.  From the middle panel in Figure
\ref{fig:supermolecule}, we find a single, statistically significant,
natural boundary for a LSE supermolecule comprising nearly all the
financial atoms in the market.  This supermolecule is predominantly
financials (atoms 1, 3, 4, 6, 7, 9, 12, 13, 14, 17, 19, many of which
are components of the LSE investment trusts molecule shown in Figure
\ref{fig:LSEmol4}) in character, and the excluded atoms --- atom 2
(oil and gas), atom 11 (foreign consumer electronics), atom 15
(foreign oil and gas), and atom 21 (restaurants) --- are all from
market sectors distant from the finance sector.

\begin{figure}[htpb]
\centering
\includegraphics[scale=0.45,clip=true]{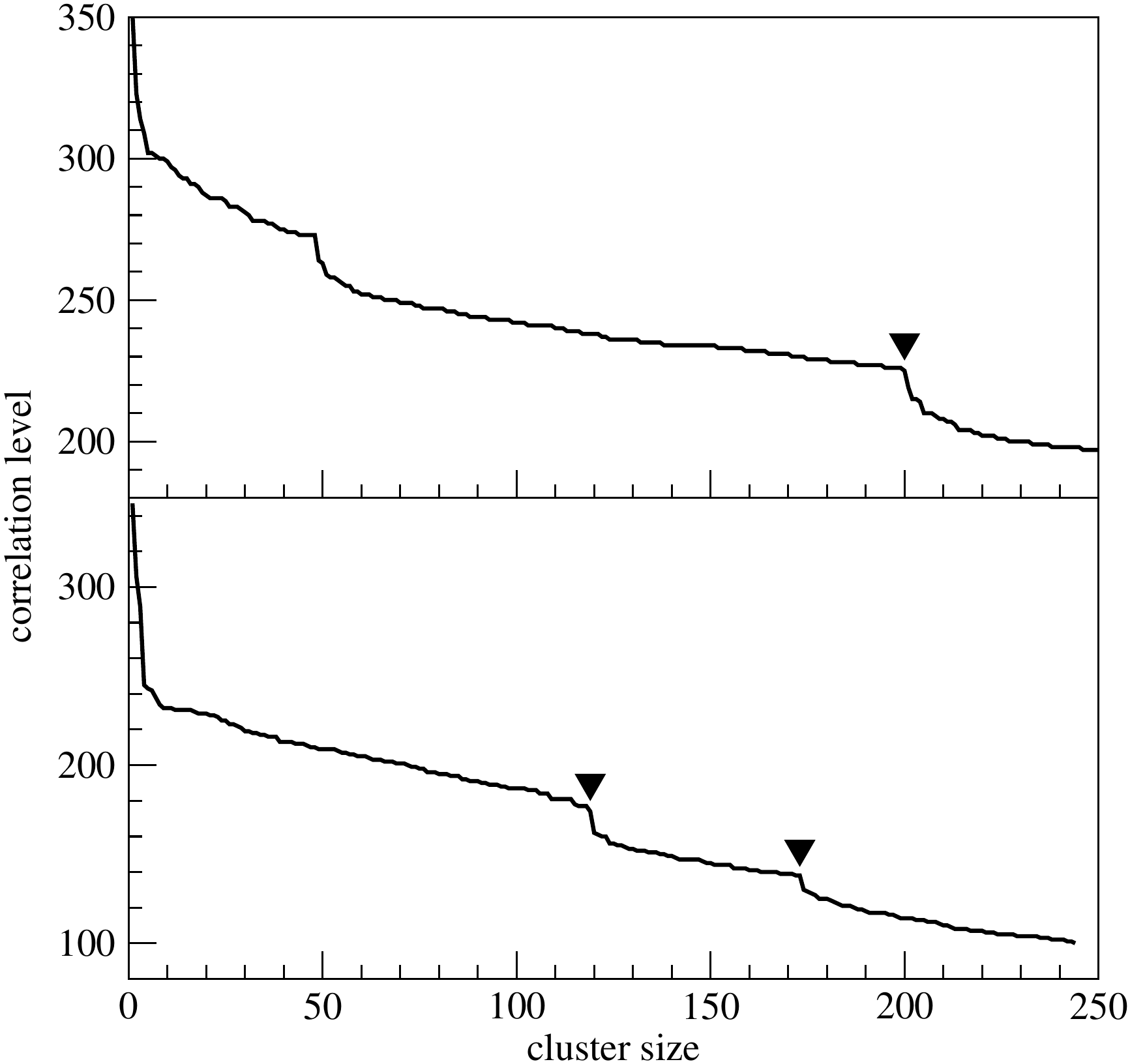}
\caption{Statistical signatures for self-organization at the financial
supermolecule level, within the complete-link partial
hierarchical clustering histories of small markets: (top) starting
from atom 7 of the HKSE, and (bottom) starting from atom 1 of the
SGX.  The natural boundaries of the financial supermolecules are
indicated by the black triangles.}
\label{fig:supermolHKSG}
\end{figure}

Of course, in the TSE, the financial molecules themselves do not
respect sector boundaries, probably because of extensive
cross-ownership practiced in this market
\cite{Roehner2005PhysicaA347p613}.  Therefore, the financial
supermolecules do not represent any particular market sector, but
represent instead some kind of \emph{`eigen'-trader}, which is a
collection of institutional and retail traders whose trading
strategies are largely mutually reinforcing within the two-year time
frame investigated.  Different `eigen'-traders probably have different
concentrations in their portfolio of stocks, and are therefore
approximately `orthogonal' in their impacts on the market.  We suspect
this `orthogonality' of trading strategies lead to financial
supermolecules being only weakly correlated, and thus an emergent
supermolecular correlation level that is significantly lower than the
molecular correlation levels.  From the bottom panel in Figure
\ref{fig:supermolecule}, we find a natural boundary for a large
financial supermolecule in the TSE.  This supermolecule incorporates a
large fraction of the 52 financial atoms found in the TSE, and the 11
financial atoms that are not incorporated are from the banks (5, 15,
34, 39, 47), insurance (2), other financing business (7), marine
transportation (11, 22), utilities (24), and atom 20, which comprises
the stocks issued by Toyota and Honda, Japan's largest automobile
makers.  The exclusion of banks from this TSE supermolecule, however,
is limited to regional banks, as atom 8, which comprises large
diversified banks, is included in the supermolecule.

In small and less stringently regulated markets like the HKSE and SGX,
where we find only a single financial molecule, market sector
boundaries are again not statistically meaningful.  For these small
markets, we again find financial supermolecules consisting of over 100
stocks.  These comprises the sole financial molecule, other financial
atoms, and various more actively traded non-atomic stocks.  The
supermolecular correlation level is lower than for the large markets,
and the separation of correlation levels is between actively-traded
stocks and illiquid stocks.  Again, this is the statistical signature
left by the `eigen'-trader, which is the effectively the whole market
in these small markets.  From Figure \ref{fig:supermolHKSG}, we find
one natural boundary for the sole HKSE financial supermolecule, and
two natural boundaries for the SGX financial supermolecule.  All
financial atoms in the HKSE and SGX are found to participate in the
respective supermolecules.  In fact, all financial atoms in the SGX
are already found to included by the first natural boundary, and so
the second natural boundary might indicate a further separation of
correlation levels caused by different liquidity levels in the market.

\section{Summary and Discussions}

In conclusion, we considered the phenomenon of separation of dynamical
time scales, which arises generically as a result of self-organization
in a complex system with a large number of interacting microscopic
variables.  We described how these separated time scales, and their
associated dynamical structures, can be learned statistically, by
examining short-time correlational statistics between the microscopic
variables over a longer time, and identifying separated correlation
levels at the long time scale.  Arguing that only robust dynamical
structures are physically meaningful, we explained why we need only
look out for separated long-time correlation levels that are
insensitive to details of the short-time and long-time statistics
examined.  After recasting the problem of finding a small number of
effective variables (the self-organized dynamical structures) from a
large number of interacting microscopic variables as a statistical
learning problem, we moved on to describe our two-time-scale
statistical clustering method.  In this method, we would first cluster
the time series of the microscopic variables on a short time scale.
Thereafter, we would construct a structurally-simple long-time
correlation matrix by counting the number of times pairs of
microscopic variables are assigned to the same short-time clusters
over the long time scale.  We then developed the method of partial
hierarchical clustering to automatically and systematically extract
the separated correlation levels, and their associated collections of
microscopic variables, from the long-time correlation matrix.

Using the two-time-scale statistical clustering method to analyze the
five equity markets shown in Table \ref{tab:markets} over the two-year
period from 2006 to 2007, we found in all markets separated
correlation levels that point to self-organization at several
hierarchical levels.  We call the separated correlation levels atomic,
molecular, and supermolecular correlation levels, and the effective
dynamical variables they are associated with financial atoms,
financial molecules, and financial supermolecules.  In all markets,
financial atoms are found to consist of around 10 stocks, having
internal correlations far above the background correlation level
caused by market-level drifts.  Some financial atoms are trivial,
consisting of two or more stocks issued by the same company, while
others have compositions expected of companies competing directly with
each other within the same industry.  Quite a number of financial
atoms, however, consist surprisingly of stocks from very different
industries.  We find that these surprising atomic compositions can
only be explained in terms of the ownership profiles and broader
business conflicts and mutual interests of the constituent stocks.  In
the more globalized NYSE and LSE, we also find foreign atoms, which
are part of financial atoms in their home markets.  

In the smaller markets (HKSE and SGX), we find a single financial
molecule made up of around 10 financial atoms.  In the larger markets
(NYSE, LSE, and TSE), we find many financial molecules, with sizes
ranging from several to around 20 atoms.  In the NYSE and LSE,
financial molecules comprise atoms from a small number of closely
related market sectors, whereas in the TSE, HKSE, and SGX, financial
molecules do not respect market sector boundaries, sporting
constituent atoms from distant sectors.  We suspect that the former
reflects the stringent market regulations in place in the NYSE and
LSE, while the latter is a consequence of extensive cross-ownership in
the TSE, and dominance by a small number of multi-industry holding
companies in the HKSE and SGX.  In the NYSE and LSE, market sector
boundaries continue to be more or less respected by their financial
supermolecules, and as such, market sectors can serve as high-level
effective variables for understanding the dynamics in these markets.
In contrast, the high-level effective variables in the TSE, HKSE, and
SGX are multi-sector financial supermolecules associated with `eigen'
collections of institutional and retail traders with roughly
`orthogonal' market impacts.  Close examinations of the correlational
structures of the NYSE, HKSE, and SGX financial molecules reveal very
interesting signatures of such large market events as the Subprime
Crisis and the Chinese Correction.  No extraordinary structures were
seen in the LSE and TSE financial molecules.  This led us to conclude
that precursor signatures of large market events can only be detected
in their markets of origin.

It is important to point out here that financial atoms and molecules
are not merely curiosity items.  The robust identification of
financial atoms and molecules forms the first step in our program to
understand real financial markets.  Once the financial atoms and
molecules are found, we will use them as effective variables to build
effective dynamical models.  Under normal market conditions, we will
then apply understanding derived from this effective modeling to risk
analysis and portfolio management.  Portfolio optimization schemes
based on the minimal spanning tree analysis of market correlations
have been developed \cite{Onnela2003PhysRevE68e056110,
Tola2008JEconDynControl32p235}, but we believe it is possible to bring
the state of the art to a whole new level.  For example, we believe
current hedging strategies used by large fund managers must be
revised, in line with the structures of financial atoms and molecules
found in the respective markets.  In particular, when designing a
portfolio, it is important to observe that fluctuations in the
price-weighted mean (the center of mass) of large financial molecules
are suppressed relative to market-level fluctuations, because of the
combined inertia of all their strongly-correlated constituent stocks.
On the other hand, we need to be more careful investing in small
financial molecules, because they are `lighter', and are thus more
susceptible to driving by market `forces'.  More importantly, by
making trading decisions at the financial molecule level, it is
possible to unambiguously decompose the risk any given portfolio is
exposed to, into a systematic risk component associated with the
effectively deterministic dynamics of the financial molecules, and a
stochastic risk component associated with random, unpredictable shocks
experienced by the market on a whole.  Our findings suggest that the
systematic risk can be further reduced by taking into account the
effective dynamics of the financial supermolecules.

Our analyses and findings have even more serious implications to
understanding and coping with financial crises.  As suggested by
previous works \cite{Onnela2003PhysicaA324p247,
Araujo2007QuantFin7p63, Bonanno2001PhysicaA299p16,
Araujo2008PhysLettA372p429}, and also judging by the intriguing
results seen in the NYSE, HKSE, and SGX, we believe a market becomes
extremely susceptible to crashes when a significant fraction of stocks
gets lock into a single financial supermolecule.  If we can determine
the critical size of such a market-spanning supermolecule, and also
continuously track the sizes of growing supermolecules, we might be
able to either forecast the onset of a financial crisis one or two
months in advance, or perhaps even avert the crisis entirely.  To this
end, we are currently looking at the Chinese Correction on a shorter
time scale.  By tracking the dynamics of financial molecules in the
SGX, and the evolution of their molecular structures from month to
month over 2006 and 2007, we aim to develop a dynamical picture of
bond formation prior to, and bond breaking after the market crash.  We
believe such a `chemical' picture will offer important clues to
understanding the end-February 2007 Chinese Correction, which probably
catalyzed the Subprime Crisis that surfaced a few months later.

\begin{acknowledgments}
This research is supported in part by the Nanyang Technological
University startup grant SUG 19/07, and also by the Nanyang
Technological University's CN Yang Scholars Programme.  We have had
helpful discussions with Low Buen Sin, Charlie Charoenwong, Gerald
Cheang Hock Lye, and Chris Kok Jun Liang.  
\end{acknowledgments}

\begin{appendix}

\section{Lists of Strong Financial Atoms}

\subsection{New York Stock Exchange}

\begin{center}
\centering\footnotesize
\begin{tabular}{ccccc}
\hline
Atom & Code & Name & Market Sector & Remarks \\
\hline
1 & VIA & Viacom A & Services & \\
  & VIA.B & Viacom B & Services & \\
\hline
2 & LEN & Lennar A & \parbox[t]{2cm}{Industrial Goods} & \parbox[t]{2cm}{Residential Construction} \\
  & LEN.B & Lennar B & \parbox[t]{2cm}{Industrial Goods} & \parbox[t]{2cm}{Residential Construction} \\
\hline
\end{tabular}
\end{center}

\begin{center}
\centering\footnotesize
\begin{tabular}{ccccc}
\hline
Atom & Code & Name & Market Sector & Remarks \\
\hline
3 & PBR & \parbox[t]{2cm}{Petroleo Brasileiro SA} &
\parbox[t]{2cm}{Basic Materials} & ADR \\
  & PBR.A & \parbox[t]{2cm}{Petroleo Brasileiro SA} & Basic
  Materials & ADR \\
\hline
4 & RDS.A & \parbox[t]{2cm}{Royal Dutch Shell A} & Basic Materials & ADR \\
  & RDS.B & \parbox[t]{2cm}{Royal Dutch Shell B} & Basic Materials & ADR \\
\hline
5 & CBS & CBS & Services & \\
  & CBS.A & CBS & Services & \\
\hline
6 & BHP & BHP Billiton & Basic Materials & ADR \\
  & BBL & BHP Billiton & Basic Materials & ADR \\
  & RTP & Rio Tinto plc & Basic Materials & ADR \\
\hline
7 & LAZ & Lazard & Financial & \parbox[t]{2cm}{Asset Management} \\
  & LDZ & Lazard 6.625\% & Financial & \parbox[t]{2cm}{Equity Security Units} \\
\hline
8 & UN & Unilever NV & Consumer Goods & ADR \\
  & UL & Unilever plc & Consumer Goods & ADR \\
\hline
9 & CCL & Carnival Corp & Services & \parbox[t]{2cm}{General
Entertainment} \\
  & CUK & Carnival plc & Services & ADR \\
  & RCL & \parbox[t]{2cm}{Royal Carribean Cruises} & Services &
  \parbox[t]{2cm}{General Entertain\-ment} \\
\hline
10 & RIO & Vale & Basic Materials & ADR \\
   & RIO- & Vale & Basic Materials & ADR \\
\hline
11 & FCE.A & \parbox[t]{2cm}{Forest City Enterprises A} & Financial & \parbox[t]{2cm}{Property Management} \\
   & FCE.B & \parbox[t]{2cm}{Forest City Enterprises B} & Financial & \parbox[t]{2cm}{Property Management} \\
\hline
12 & NWS & News Corp & Services & \\
   & NWS.A & News Corp A & Services & \\
\hline
13 & ETR & Entergy Corp & Utilities & \parbox[t]{2cm}{Electric
Utilities} \\
   & ETR.PRA & \parbox[t]{2cm}{Entergy Corp 7.625\%} & Utilities & \parbox[t]{2cm}{Equity
	Security Units} \\
\hline
14 & ENL & \parbox[t]{2cm}{Reed Elsevier NV} & Services & ADR \\
   & RUK & \parbox[t]{2cm}{Reed Elsevier plc} & Services & ADR \\
\hline
15 & BBD & \parbox[t]{2cm}{Banco Bradesco SA} & Financial & ADR \\
   & ITU & \parbox[t]{2cm}{Banco Itau Hldg Financeira SA} & Financial & ADR \\
\hline
16 & DDR & \parbox[t]{2cm}{Developers Diversified Realty Corp} & Financial & Retail
REIT \\
   & REG & \parbox[t]{2cm}{Regency Centers Corp} & Financial & Retail REIT \\
	& MAC & \parbox[t]{2cm}{The Macerich Co} & Financial & Retail REIT \\
	& SPG & \parbox[t]{2cm}{Simon Property Gp Inc} & Financial & Retail REIT \\
	& EQR & \parbox[t]{2cm}{Equity Residential} & Financial & \parbox[t]{2cm}{Residential REIT} \\
	& GGP & \parbox[t]{2cm}{General Growth Properties Inc} & Financial & \parbox[t]{2cm}{Retail REIT} \\
\hline
\end{tabular}
\end{center}

\begin{center}
\centering\footnotesize
\begin{tabular}{ccccc}
\hline
Atom & Code & Name & Market Sector & Remarks \\
\hline
17 & KBH & KB Home & \parbox[t]{2cm}{Industrial Goods} & \parbox[t]{2cm}{Residential Construction} \\
   & PHM & Pulte Homes Inc & \parbox[t]{2cm}{Industrial Goods} & \parbox[t]{2cm}{Residential Construction} \\
	& CTX & Centex Corp & \parbox[t]{2cm}{Industrial Goods} & \parbox[t]{2cm}{Residential Construction} \\
	& MDC & MDC Hldg Inc & \parbox[t]{2cm}{Industrial Goods} & \parbox[t]{2cm}{Residential Construction} \\
	& TOL & Toll Brothers Inc & \parbox[t]{2cm}{Industrial Goods} & \parbox[t]{2cm}{Residential Construction} \\
	& DHI & DR Horton Inc & \parbox[t]{2cm}{Industrial Goods} & \parbox[t]{2cm}{Residential Construction} \\
	& RYL & \parbox[t]{2cm}{The Ryland Gp Inc} & \parbox[t]{2cm}{Industrial Goods} & \parbox[t]{2cm}{Residential Construction} \\
	& HOV & \parbox[t]{2cm}{Hovnanian Enterprises Inc} & \parbox[t]{2cm}{Industrial Goods} & \parbox[t]{2cm}{Residential Construction} \\
	& SPF & \parbox[t]{2cm}{Standard Pacific Corp} & \parbox[t]{2cm}{Industrial Goods} & \parbox[t]{2cm}{Residential Construction} \\
	& MTH & \parbox[t]{2cm}{Meritage Homes Corp} & \parbox[t]{2cm}{Industrial Goods} & \parbox[t]{2cm}{Residential Construction} \\
	& BHS & \parbox[t]{2cm}{Brookfield Homes Corp} & \parbox[t]{2cm}{Industrial Goods} & \parbox[t]{2cm}{Residential Construction} \\
\hline
18 & E & Eni SPA & Basic Materials & ADR \\
   & TOT & Total SA & Basic Materials & ADR \\
	& BP & BP plc & Basic Materials & ADR \\
	& REP & \parbox[t]{2cm}{Repsol YPF SA} & Basic Materials & ADR \\
\hline
19 & MET & MetLife Inc & Financial & \parbox[t]{2cm}{Life Assurance} \\
   &     & \parbox[t]{2cm}{MetLife 6.375\%} & Financial & \parbox[t]{2cm}{Equity Security
	Units} \\
\hline
20 & ABV & \parbox[t]{2cm}{Companhia de Bebidas das Americas} & \parbox[t]{2cm}{Consumer Goods} & ADR \\
   & ABV.C & \parbox[t]{2cm}{Companhia de Bebidas das Americas} & \parbox[t]{2cm}{Consumer Goods} & ADR \\
\hline
21 & CVX & \parbox[t]{2cm}{Chevron Corp} & Basic Materials & \parbox[t]{2cm}{Integrated Oil \& Gas} \\
   & XOM & \parbox[t]{2cm}{Exxon Mobile Corp} & Basic Materials & \parbox[t]{2cm}{Integrated Oil \& Gas} \\
	& PEO & \parbox[t]{2cm}{Petroleum \& Resources Corp} & Financial & \parbox[t]{2cm}{Equity Closed-End Fund} \\
	& COP & \parbox[t]{2cm}{ConocoPhillips} & Basic Materials & \parbox[t]{2cm}{Integrated Oil \& Gas} \\
	& OXY & \parbox[t]{2cm}{Occidental Petroleum Corp} & Basic Materials & \parbox[t]{2cm}{Independent Oil \& Gas} \\
	& MRO & \parbox[t]{2cm}{Marathon Oil Corp} & Basic Materials & \parbox[t]{2cm}{Oil \& Gas Refining \& Marketing} \\
\hline
22 & ESV & Ensco Intl Inc & Basic Materials & \parbox[t]{2cm}{Oil \& Gas Drilling \& Exploration} \\
	& NE  & Noble Corp & Basic Materials & \parbox[t]{2cm}{Oil \& Gas Drilling \& Exploration} \\
	& PDE & Pride Intl Inc & Basic Materials & \parbox[t]{2cm}{Oil \& Gas Drilling \& Exploration} \\
\hline
23 & BRP & \parbox[t]{2cm}{Brasil Telecom Participacoes SA} & Technology & ADR \\
	& BTM & \parbox[t]{2cm}{Brasil Telecom SA} & Technology & ADR \\
	& TBH & \parbox[t]{2cm}{Brazilian Telecom HOLDRS} & Technology & ADR \\
\hline
24 & SLB & Schlumberger & Basic Materials & \parbox[t]{2cm}{Oil \& Gas Equipment \& Services} \\
	& SII & Smith Intl Inc & Basic Materials & \parbox[t]{2cm}{Oil \& Gas Equipment \& Services} \\
\hline
\end{tabular}
\end{center}

\begin{center}
\centering\footnotesize
\begin{tabular}{ccccc}
\hline
Atom & Code & Name & Market Sector & Remarks \\
\hline
25 & KV.A & KV Pharma A & Healthcare & \parbox[t]{2cm}{Drug Delivery} \\
	& KV.B & KV Pharma B & Healthcare & \parbox[t]{2cm}{Drug Delivery} \\
\hline
26 & CNQ & \parbox[t]{2cm}{Canadian Natural Resources} & Basic Materials & \parbox[t]{2cm}{Independent Oil \& Gas} \\
	& SU  & \parbox[t]{2cm}{Suncor Energy Inc} & Basic Materials & \parbox[t]{2cm}{Independent Oil \& Gas} \\
	& ECA & \parbox[t]{2cm}{EnCana Corp} & Basic Materials & \parbox[t]{2cm}{Independent Oil \& Gas} \\
	& TLM & \parbox[t]{2cm}{Talisman Energy Inc} & Basic Materials & \parbox[t]{2cm}{Independent Oil \& Gas} \\
	& SFY & \parbox[t]{2cm}{Swift Energy Co} & Basic Materials & \parbox[t]{2cm}{Independent Oil \& Gas} \\
	& CHK & \parbox[t]{2cm}{Chesapeake Energy Corp} & Basic Materials & \parbox[t]{2cm}{Independent Oil \& Gas} \\
	& NXY & Nexen Inc & Basic Materials & \parbox[t]{2cm}{Independent Oil \& Gas} \\
	& EOG & \parbox[t]{2cm}{EOG Resources Inc} & Basic Materials & \parbox[t]{2cm}{Independent Oil \& Gas} \\
	& PCZ & Petro-Canada & Basic Materials & \parbox[t]{2cm}{Oil \& Gas Refining \& Marketing} \\
\hline
27 & MIR & Mirant Corp & Utilities & \\
	& MIR.WSA & Mirant Corp & Utilities & \\
	& MIR.WSB & Mirant Corp & Utilities & \\
\hline
28 & HEI & HEICO Corp & \parbox[t]{2cm}{Industrial Goods} & \parbox[t]{2cm}{Aerospace \& Defense} \\
	& HEI.A & HEICO Corp & \parbox[t]{2cm}{Industrial Goods} & \parbox[t]{2cm}{Aerospace \& Defense} \\
\hline
29 & BBI & \parbox[t]{2cm}{Blockbuster Inc A} & Services & \\
	& BBI.B & \parbox[t]{2cm}{Blockbuster Inc B} & Services & \\
\hline
30 & BNI & \parbox[t]{2cm}{Burlington Northern Santa Fe Corp} & Services & Railroads \\
	& UNP & \parbox[t]{2cm}{Union Pacific Corp} & Services & Railroads \\
	& CSX & CSX Corp & Services & Railroads \\
	& NSC & \parbox[t]{2cm}{Norfolk Southern Corp} & Services & Railroads \\
\hline
\end{tabular}
\end{center}

\subsection{London Stock Exchange}

\begin{center}
\centering\footnotesize
\begin{tabular}{ccccc}
\hline
Atom & Code & Name & Market Sector & Remarks \\
\hline
1 & SDR & Schroders & Financials & \parbox[t]{2cm}{Investment Services} \\
  & SDRt & Schroders & Financials & \parbox[t]{2cm}{Investment Services} \\
\hline
2 & RDSa & \parbox[t]{2cm}{Royal Dutch Shell} & Energy & \parbox[t]{2cm}{Integrated Oil \& Gas} \\
  & RDSb & \parbox[t]{2cm}{Royal Dutch Shell} & Energy & \parbox[t]{2cm}{Integrated Oil \& Gas} \\
\hline
3 & HMSO & Hammerson & Financials & REIT \\
  & LII & Liberty Intl & Financials & REIT \\
  & LAND & \parbox[t]{2cm}{Land Securities Gp} & Financials & REIT \\
  & BXTN & Brixton & Financials & \parbox[t]{2cm}{Real Estate Ops} \\
\hline
4 & SCIN & \parbox[t]{2cm}{Scottish Investment Trust} & Financials & \parbox[t]{2cm}{Investment Trusts} \\
  & WTAN & \parbox[t]{2cm}{Witan Investment Trust} & Financials & \parbox[t]{2cm}{Investment Trusts} \\
\hline
5 & AAL & \parbox[t]{2cm}{Anglo American} & Basic Materials & \parbox[t]{2cm}{Mining \& Metals} \\
  & BLT & \parbox[t]{2cm}{BHP Billiton} & Basic Materials & \parbox[t]{2cm}{Mining \& Metals} \\
  & RIO & \parbox[t]{2cm}{Rio Tinto} & Basic Materials & \parbox[t]{2cm}{Mining \& Metals} \\
\hline
\end{tabular}
\end{center}

\begin{center}
\centering\footnotesize
\begin{tabular}{ccccc}
\hline
Atom & Code & Name & Market Sector & Remarks \\
\hline
6 & MNKS & \parbox[t]{2cm}{Monks Investment Trust} & Financials & \parbox[t]{2cm}{Investment Trusts} \\
  & SMT & \parbox[t]{2cm}{Scottish Mortgage Investment Trust} & Financials & \parbox[t]{2cm}{Investment Trusts} \\
\hline
7 & EDIN & \parbox[t]{2cm}{Edinburgh Investment Trust} & Financials & \parbox[t]{2cm}{Investment Trusts} \\
  & MRCM & \parbox[t]{2cm}{Mercantile Investment Trust} & Financials & \parbox[t]{2cm}{Investment Trusts} \\
\hline
8 & ANTO & \parbox[t]{2cm}{Antofagasta} & Basic Materials & \parbox[t]{2cm}{Mining \& Metals} \\
  & VED & \parbox[t]{2cm}{Vedanta Resources} & Basic Materials & \parbox[t]{2cm}{Mining \& Metals} \\
\hline
9 & ATST & \parbox[t]{2cm}{Alliance Trust} & Financials & \parbox[t]{2cm}{Investment Trusts} \\
  & FEV & \parbox[t]{2cm}{Fidelity European Values} & Financials & \parbox[t]{2cm}{Investment Trusts} \\
  & RCP & \parbox[t]{2cm}{RIT Capital Partners} & Financial & \parbox[t]{2cm}{Investment Trusts} \\
\hline
10 & BDEV & \parbox[t]{2cm}{Barratt Developments} & \parbox[t]{2cm}{Cyclical Consumer Goods \& Services} & Homebuilding \\
   & PSN & Persimmon & \parbox[t]{2cm}{Cyclical Consumer Goods \& Services} & Homebuilding \\
	& RDW & Redrow & \parbox[t]{2cm}{Cyclical Consumer Goods \&
	Services} & Homebuilding \\
\hline
11 & 0593q & \parbox[t]{2cm}{Samsung Electronics} & \parbox[t]{2cm}{Cyclical Consumer Goods \& Services} & \parbox[t]{2cm}{Consumer Electronics} \\
   & 0593xq & \parbox[t]{2cm}{Samsung Electronics} & \parbox[t]{2cm}{Cyclical Consumer Goods \& Services} & \parbox[t]{2cm}{Consumer Electronics} \\
\hline
12 & AV & Aviva & Financials & \parbox[t]{2cm}{Insurance - Multiline} \\
   & PRU & Prudential & Financials & \parbox[t]{2cm}{Insurance - Life \& Health} \\
	& LGEN & \parbox[t]{2cm}{Legal and General Gp} & Financials & \parbox[t]{2cm}{Insurance - Multiline} \\
	& FP & \parbox[t]{2cm}{Friends Provident} & Financials & \parbox[t]{2cm}{Insurance - Life \& Health} \\
	& STAN & \parbox[t]{2cm}{Standard Chartered} & Financials & Banks \\
	& FRCL & \parbox[t]{2cm}{Foreign \& Colonial Investment Trust} & Financials & \parbox[t]{2cm}{Investment Trusts} \\
\hline
\end{tabular}
\end{center}

\subsection{Tokyo Stock Exchange}

\begin{center}
\centering\footnotesize
\begin{tabular}{ccccl}
\hline
Atom & Code & Name & Market Sector & Remarks \\
\hline
1 & 8001 & Itochu Corp & \parbox[t]{2cm}{Wholesale Trade} & Section 1 \\
& 8002 & \parbox[t]{2cm}{Marubeni Corp} & \parbox[t]{2cm}{Wholesale Trade} & Section 1 \\
& 8031 & \parbox[t]{2cm}{Mitsui \& Co} & \parbox[t]{2cm}{Wholesale Trade} & Section 1 \\
& 8053 & \parbox[t]{2cm}{Sumitomo Corp} & \parbox[t]{2cm}{Wholesale Trade} & Section 1 \\
& 8058 & \parbox[t]{2cm}{Mitsubishi Corp} & \parbox[t]{2cm}{Wholesale Trade} & Section 1 \\
\hline
\end{tabular}
\end{center}

\begin{center}
\centering\footnotesize
\begin{tabular}{ccccl}
\hline
Atom & Code & Name & Market Sector & Remarks \\
\hline
2 & 8725 & \parbox[t]{2cm}{Mitsui Sumitomo Insurance Gp Hldg} & Insurance & Section 1 \\
& 8754 & \parbox[t]{2cm}{NIPPONKOA Insurance Co.} & Insurance & Section 1 \\
& 8755 & \parbox[t]{2cm}{Sompo Japan Insurance Inc.} & Insurance & Section 1 \\
& 8759 & \parbox[t]{2cm}{Nissay Dowa General Insurance Co.} & Insurance & Section 1 \\
& 8761 & \parbox[t]{2cm}{Aioi Insurance Co} & Insurance & Section 1 \\
& 8763 & \parbox[t]{2cm}{Fuji Fire \& Marine Insurance Co} & Insurance & Section 1 \\
\hline
3 & 8601 & \parbox[t]{2cm}{Daiwa Securities Gp Inc} & Securities & Section 1 \\
& 8604 & \parbox[t]{2cm}{Nomura Hldg Inc} & Securities & Section 1 \\
& 8606 & \parbox[t]{2cm}{Shinko Securities Co} & Securities & Section 1 \\
& 8607 & \parbox[t]{2cm}{Mizuho Investors Securities Co} & Securities & Section 1 \\
& 8609 & \parbox[t]{2cm}{Okasan Securities Gp Inc} & Securities & Section 1 \\
& 8613 & \parbox[t]{2cm}{Marusan Securities Co} & Securities & Section 1 \\
& 8614 & \parbox[t]{2cm}{Toyo Securities Co} & Securities & Section 1 \\
& 8616 & \parbox[t]{2cm}{Tokai Tokyo Securities Co} & Securities & Section 1 \\
& 8622 & \parbox[t]{2cm}{Mito Securities Co} & Securities & Section 1 \\
& 8624 & \parbox[t]{2cm}{Ichiyoshi Securities Co} & Securities & Section 1 \\
\hline
4 & 8801 & \parbox[t]{2cm}{Mitsui Fudosan Co} & Real Estate & Section 1 \\
& 8802 & \parbox[t]{2cm}{Mitsubishi Estate Co} & Real Estate & Section 1 \\
& 8815 & \parbox[t]{2cm}{Tokyu Land Co} & Real Estate & Section 1 \\
& 8830 & \parbox[t]{2cm}{Sumitomo Realty \& Development Co} & Real
Estate & Section 1 \\
\hline
5 & 8356 & \parbox[t]{2cm}{The Juroku Bank} & Banks & Section 1 \\
& 8360 & \parbox[t]{2cm}{The Yamanashi Chuo Bank} & Banks & Section 1 \\
& 8362 & The Fukui Bank & Banks & Section 1 \\
& 8368 & \parbox[t]{2cm}{The Hyakugo Bank} & Banks & Section 1 \\
& 8381 & \parbox[t]{2cm}{The San-in Godo Bank} & Banks & Section 1 \\
& 8382 & \parbox[t]{2cm}{The Chugoku Bank} & Banks & Section 1 \\
& 8386 & \parbox[t]{2cm}{The Hyakujushi Bank} & Banks & Section 1 \\
& 8388 & The Awa Bank & Banks & Section 1 \\
& 8390 & \parbox[t]{2cm}{The Kagoshima Bank} & Banks & Section 1 \\
& 8394 & The Higo Bank & Banks & Section 1 \\
\hline
6 & 5401 & \parbox[t]{2cm}{Nippon Steel Corp} & Steel \& Iron & Section 1 \\
& 5405 & \parbox[t]{2cm}{Sumitomo Metal Industries} & Steel \& Iron & Section 1 \\
& 5406 & Kobe Steel & Steel \& Iron & Section 1 \\
& 5411 & JFE Hldg Inc & Steel \& Iron & Section 1 \\
\hline
\end{tabular}
\end{center}

\begin{center}
\centering\footnotesize
\begin{tabular}{ccccl}
\hline
Atom & Code & Name & Market Sector & Remarks \\
\hline
7 & 8253 & \parbox[t]{2cm}{Credit Saison Co} & \parbox[t]{2cm}{Other Financing
Business} & Section 1 \\
& 8515 & Aiful Corp & \parbox[t]{2cm}{Other Financing
Business} & Section 1 \\
& 8564 & Takefuji Corp & \parbox[t]{2cm}{Other Financing
Business} & Section 1 \\
& 8572 & Acom Co & \parbox[t]{2cm}{Other Financing
Business} & Section 1 \\
& 8574 & Promise Co & \parbox[t]{2cm}{Other Financing
Business} & Section 1 \\
& 8583 & \parbox[t]{2cm}{Mitsubishi UFJ Financial Gp Inc} & 
\parbox[t]{2cm}{Other Financing Business} & Section 1 \\
\hline
8 & 8308 & \parbox[t]{2cm}{Resona Hldg Inc} & Banks & Section 1 \\
& 8309 & \parbox[t]{2cm}{Chuo Mitsui Trust Hldg Inc} & Banks & Section 1 \\
& 8403 & \parbox[t]{2cm}{The Sumitomo Trust \& Banking Co} & Banks &
Section 1 \\
& 8404 & \parbox[t]{2cm}{Mizuho Trust \& Banking Co} & Banks &
Section 1 \\
& 8411 & \parbox[t]{2cm}{Mizuho Financial Gp Inc} & Banks & Section 1 \\
\hline
9 & 4321 & Kenedix Inc & Services & Section 1 \\
& 8804 & \parbox[t]{2cm}{Tokyo Tatemono Co} & Real Estate & Section 1 \\
& 8868 & Urban\textsuperscript{*} & Real Estate & Section 1 \\
& 8888 & Creed Co & Real Estate & Section 1 \\
& 8902 & \parbox[t]{2cm}{Pacific Hldg Inc} & Real Estate & Section 1 \\
\hline
10 & 5726 & \parbox[t]{2cm}{Osaka Titanium Technologies Co} &
\parbox[t]{2cm}{Non-Ferrous Metals} & Section 1 \\
& 5727 & \parbox[t]{2cm}{Toho Titanium Co} &
\parbox[t]{2cm}{Non-Ferrous Metals} & Section 1 \\
\hline
11 & 9104 & \parbox[t]{2cm}{Mitsui O.S.K. Lines} & \parbox[t]{2cm}{Marine
Transportation} & Section 1 \\
& 9107 & \parbox[t]{2cm}{Kawasaki Kisen Kaisha} &
\parbox[t]{2cm}{Marine Transportation} & Section 1 \\
\hline
12 & 1801 & Taisei Corp & Construction & Section 1 \\
& 1802 & Obayashi Corp & Construction & Section 1 \\
& 1803 & Shimizu Corp & Construction & Section 1 \\
& 1812 & Kajima Corp & Construction & Section 1 \\
\hline
13 & 5711 & \parbox[t]{2cm}{Mitsubishi Materials Corp} & \parbox[t]{2cm}{Non-Ferrous Metals} & Section 1 \\
& 5713 & \parbox[t]{2cm}{Sumitomo Metal Mining Co} &
\parbox[t]{2cm}{Non-Ferrous Metals} & Section 1 \\
& 5771 & Mitsubishi Shind & - & - \\
\hline
14 & 4751 & CyberAgent Inc & - & Mothers \\
& 4788 & \parbox[t]{2cm}{Cyber Communications Inc} & - & Mothers \\
\hline
\end{tabular}
\vskip .5\baselineskip
\textsuperscript{*}This instrument code no longer exist on the TSE,
but may be related to Urban Life Co, in the real estate sales,
brokerage, and leasing business, traded on the Osaka Securities
Exchange as 8851.
\end{center}

\subsection{Hong Kong Stock Exchange}

\begin{center}
\centering\footnotesize
\begin{tabular}{ccccl}
\hline
Atom & Code & Name & Market Sector & Remarks \\
\hline
1 & 0001 & \parbox[t]{2cm}{Cheung Kong Hldg} & Properties \\
& 0011 & Hang Seng Bank & Finance \\
& 0013 & \parbox[t]{2cm}{Hutchison Whampoa} & \parbox[t]{2cm}{Consolidated
Enterprises} \\
& 0016 & \parbox[t]{2cm}{Sun Hung Kai Properties} & Properties \\
& 0066 & MTR Corp & Utilities \\
& 0941 & China Mobile & \parbox[t]{2cm}{Consolidated Enterprises} \\
\hline
2 & 0762 & \parbox[t]{2cm}{China Unicom (HK)} & \parbox[t]{2cm}{Consolidated Enterprises} \\
& 2318 & \parbox[t]{2cm}{Ping An Insurance (Gp) Co of China} & Finance \\
& 2628 & \parbox[t]{2cm}{China Life Insurance Co} & Finance \\
\hline
3 & 0688 & \parbox[t]{2cm}{China Overseas Land \& Investment} &
Properties \\
& 1109 & \parbox[t]{2cm}{China Resources Land} & Properties \\
& 3383 & \parbox[t]{2cm}{Agile Properties Hldg} & Properties \\
\hline
4 & 2899 & \parbox[t]{2cm}{Zijin Mining Gp Co} & Miscellaneous \\
& 3330 & \parbox[t]{2cm}{Lingbao Gold Co} & Miscellaneous \\
\hline
5 & 0323 & \parbox[t]{2cm}{Maanshan Iron \& Steel Co} & Industries \\
& 0347 & \parbox[t]{2cm}{Angang Steel Co} & Industries \\
\hline
6 & 0019 & Swire Pacific A & \parbox[t]{2cm}{Consolidated
Enterprises} \\
& 0087 & Swire Pacific B & \parbox[t]{2cm}{Consolidated Enterprises} \\
\hline
7 & 0857 & PetroChina Co & Industries & \parbox[t]{2cm}{Oil \& Natural
Gas} \\
& 0883 & CNOOC & Industries & \parbox[t]{2cm}{Oil \& Natural Gas} \\
\hline
\end{tabular}
\end{center}

\subsection{Singapore Stock Exchange}

\begin{center}
\centering\footnotesize
\begin{tabular}{ccccl}
\hline
Atom & Code & Name & Market Sector & Remarks \\
\hline
1 & Z74 & Singtel & TSC & \\
& & Singtel 10 & TSC & \\
& & Singtel 100 & TSC & \\
\hline
2 & C6L & \parbox[t]{2cm}{Singapore Airlines} & TSC & \\
& & \parbox[t]{2cm}{Singapore Airlines 200} & TSC & \\
\hline
3 & C56 & \parbox[t]{2cm}{Celestial Nutrifoods} & Manufacturing \\
& C86 & \parbox[t]{2cm}{China Sun Bio-Chem Tech Gp Co} &
Manufacturing \\
\hline
4 & C68 & Mirach Energy & Services \\
& W81 & \parbox[t]{2cm}{Sky China Petroleum Svcs} & Services \\
& F33 & Ferrochina & Manufacturing \\
& E90 & \parbox[t]{2cm}{China Sky Chemical Fibre Co} & Manufacturing
\\
\hline
5 & C31 & CapitaLand & Properties \\
& C09 & City Development & Properties \\
\hline
6 & D05 & DBS Gp Hldg & Finance \\
& U11 & \parbox[t]{2cm}{United Overseas Bank} & Finance \\
& O39 & \parbox[t]{2cm}{Oversea-Chinese Banking Corp} & Finance \\
& W05 & Wing Tai Hldg & Properties \\
\hline
\end{tabular}
\end{center}

\end{appendix}


\begin{thebibliography}{99}
\bibitem{Transactions} All transactions through electronic order books
are automatically logged, and subsequently incorporated into various
high-frequency financial databases.  Some over-the-counter (OTC)
transactions are also logged, but these do not always make their way
into the main databases.
\bibitem{Significance} In principle, we can sacrifice some statistical
significance, and use a shorter long-time window for better temporal
resolution.  Alternatively, we can work with higher-frequency time
series data.
\bibitem{C1C2} This can be done in many ways, but we find identifying
$c_1$ with the average of the maximum intra-atom correlation and $c_2$
with the average of mininum intra-atom correlation from all
constituent atoms works most of the time.
\end{thebibliography}
\end{document}